\documentclass[fleqn,usenatbib]{mnras}

\usepackage[T1]{fontenc}
\usepackage{ae,aecompl}

\usepackage{graphicx}	
\usepackage{savesym}
\usepackage{amsmath}    
\savesymbol{iint}
\usepackage{txfonts}
\restoresymbol{TXF}{iint}
\usepackage{amssymb}	

\title[HFQPOs in MHD, eccentric discs]
{HFQPOs and discoseismic mode excitation in eccentric, relativistic discs. II. Magnetohydrodynamic simulations}

\author[J. Dewberry et al.]{
Janosz W. Dewberry,$^{1,2,3}$\thanks{E-mail: janosz.dewberry@cantab.net}
Henrik N. Latter,$^{3}$
Gordon I. Ogilvie,$^{3}$
Sebastien Fromang$^{4,5}$
\\
$^{1}$Tsung-Dao Lee Institute, No. 800 Dongchuan Road, Minhang District, Shanghai 200240, China\\
$^{2}$Department of Astronomy, Center for Astrophysics and Planetary Science, Cornell University, Ithaca, NY 14853, USA\\
$^{3}$DAMTP, University of Cambridge, CMS, Wilberforce Road, Cambridge, CB3 0WA, UK\\
$^{4}$Laboratoire AIM, CEA/DSM-CNRS-Universit\'e Paris 7, Irfu/Departement d'Astrophysique, CEA-Saclay, F-91191 Gif-sur-Yvette, France\\
$^{5}$Laboratoire des Sciences du Climat et de l'Environnement, LSCE/IPSL, CEA-CNRS-UVSQ, Universit\'e Paris-Saclay, F-91190 Gif-sur-Yvette, France
}

\date{Accepted XXX. Received YYY; in original form ZZZ}

\pubyear{2020}

\begin{document}
\label{firstpage}
\pagerange{\pageref{firstpage}--\pageref{lastpage}}
\maketitle

\begin{abstract}
Trapped inertial oscillations (r-modes) provide a promising explanation for high-frequency quasi-periodic oscillations (HFQPOs) observed in the emission from black hole X-ray binary systems. An eccentricity (or warp) can excite r-modes to large amplitudes, but concurrently the oscillations are likely damped by magnetohydrodynamic (MHD) turbulence driven by the magnetorotational instability (MRI). We force eccentricity in global, unstratified, zero-net-flux MHD simulations of relativistic accretion discs, and find that a sufficiently strong disc distortion generates trapped inertial waves despite this damping. In our simulations, eccentricities above $\sim 0.03$ in the inner disc excite trapped waves. In addition to the competition between r-mode damping and driving, we observe that larger amplitude eccentric structures modify and in some cases suppress MRI turbulence. Given the variety of distortions (warps as well as eccentricities) capable of amplifying r-modes, the robustness of trapped inertial wave excitation in the face of MRI turbulence in our simulations provides support for a discoseismic explanation for HFQPOs.
\end{abstract}




\section{Introduction}
In addition to outbursts, black hole X-ray binaries (BHBs) exhibit quasi-periodic oscillations (QPOs) with frequencies both high ($\sim50-450$Hz) and low ($\sim0.1-30$Hz). The high-frequency QPOs (HFQPOs) appear during `steep power law' (SPL) or `very high'  states of enhanced flux, and are imprinted on the high-energy emission associated with plasma in a hot corona. They generate particular interest because their frequencies appear to depend on the intrinsic properties of the central black hole, such as its mass. Given this dependence, a robust model for HFQPOs might provide a valuable measure of black hole spin \citep[for general reviews, see, e.g.,][]{Rem06,Done07,Mot16}.

Unfortunately, such a model remains elusive. Many potential explanations have been offered, appealing to relativistic precession \citep{Stel98,Stel99,Mot14}, resonances between the characteristic frequencies of particle oscillations \citep{Kluz01,Abr01}, the oscillations of geometrically thick fluid accretion tori \citep{Rez03,Blaes06,Frag16}, and the `discoseismic' oscillations of relativistic thin fluid discs \citep{Oka87,now91,now92,kat01}. However, all of the models offered thus far face several difficulties. For example, they must contend with uncertainties related to both the geometry of the accretion flow during the very high state \citep[e.g.,][]{Nay0}, and to the transmission of variability to the non-thermal X-ray emission in which HFQPOs are observed \citep[for a discussion, see ][]{dew19,dew20a}.

Nevertheless, the discoseismic model has distinct advantages over its competitors. In addition to producing an innermost stable circular orbit (ISCO), the non-monotonic variation of the horizontal epicyclic frequency ($\kappa$) close to a black hole introduces trapping regions for inertial and inertial-acoustic waves. Axisymmetric trapped inertial waves (frequently named g-modes, but here referred to as r-modes) provide a particularly attractive explanation for HFQPOs, since in hydrodynamic models their annular trapping cavity (defined by the maximum in $\kappa$) is independent of the plunging region within the ISCO, and their frequencies relate directly to $\max[\kappa]$ \citep[which is in turn determined primarily by the mass and spin of the central black hole in thin discs; ][]{Oka87}.

Furthermore, excitation via a non-linear coupling with warps or eccentricities in the disc provides a route to r-mode amplification \citep{kat04,kat08}. Large accretion rates associated with the very high state may aid the propagation of such distortions to the inner regions of the black hole accretion disc, where r-modes are trapped \citep{fer09}. The excitation mechanism can be described as a three-mode coupling in which a trapped axisymmetric r-mode grows in amplitude by interacting with the eccentricity or warp, and a second, non-axisymmetric inertial wave. In \cite{dew20a} (hereafter Paper I), we numerically demonstrated, with hydrodynamic simulations of eccentric discs, this amplification and its subsequent saturation. Paper I verifies and expands on previous semi-analytical calculations \citep{fer08,okt10}. 

The simulations described in Paper I demonstrate that in a hydrodynamic approximation, the saturation of r-mode growth can lead to the non-linear excitation of higher-frequency, vertically unstructured inertial-acoustic waves (sometimes called p-modes, but here referred to as f-modes to differentiate them from purely acoustic oscillations). The presence of such a secondary coupling supports the idea that the non-linear interaction of multiple discoseismic oscillations might be responsible for the appearance of multiple HFQPOs in some sources \citep[e.g.,][]{ort+14}. Additionally, the simulations in Paper I suggest that r-modes can redistribute angular momentum locally, reshaping their own trapping regions and blurring their own frequencies over time. 

However, magnetic fields and magnetohydrodynamic (MHD) turbulence complicate the picture. Not only do large-scale, coherent fields alter the geometry of the r-mode trapping region \citep{FL09,dew18,dew19}, but previous numerical investigations of MHD-turbulent relativistic discs failed to uncover significant r-mode signatures \citep{arr06,ReM09}. These numerical investigations did not include an explicit forcing mechanism for the oscillations, though, and therefore could only establish that turbulence driven by the magnetorotational instability (MRI) does not actively excite them \citep[in fact, r-modes are more likely to be damped by turbulence; ][]{lat06}. 

In this paper we seek to answer the question of whether sufficient excitation by disc eccentricity might overcome r-mode damping by turbulent fluctuations, generalizing the hydrodynamic simulations described in Paper I to include magnetic fields and MHD turbulence. We employ a vertically unstratified, cylindrical framework, and use the outer radial boundary condition described in Paper I to force eccentricity into discs with pre-saturated MRI turbulence. When streamlines become sufficiently non-circular, we find that r-mode signatures appear in the power spectral density (PSD), with frequencies enhanced by an additional restoring force from magnetic tension. These signatures reveal for the first time the excitation of trapped inertial waves in the presence of turbulence driven by the MRI, and support their viability as an explanation for HFQPOs. Inertial-acoustic wave excitation again accompanies trapped inertial wave excitation in our simulations. While the non-linear mode coupling seen in Paper I is less apparent, the additional excitation of f-modes over a range of frequencies suggests that both non-linear interactions and mode mixing are at play.

Although we focus on intermediate regimes in which the oscillations coexist with turbulent fluctuations, their excitation is further enhanced by a weakening of the MRI in our eccentric discs: we find that strongly forced eccentricities suppress the Maxwell stress and evacuate magnetic fields. We speculate that the MRI suppression is related to the intense periodic compressions and rarefactions that fluid blobs experience due to the steepening of the eccentric wave near the ISCO \citep[such steepening is predicted by non-linear analysis; see ][]{ogi19,lyn19}. Eccentricity gradients and twists generically produce variations in density \citep[e.g.,][]{ogi01}, although the compression may be less significant far from the black hole. The weakening of the MHD turbulence in our simulations sets the basis for exciting new work exploring the effects of strong disc distortion on the MRI in other environments, not least 
in black hole accretion discs formed in tidal disruption events (TDEs).

We describe our numerical setup in Section \ref{sec:num}, present our simulation results in Section \ref{sec:res} and provide discussion and conclusions in Sections \mbox{\ref{sec:disc}} and \mbox{\ref{sec:conc}}. For theoretical background on discoseismology and the r-mode excitation mechanism, we refer readers to \mbox{\cite{kat01}}, \mbox{\cite{dew19}}, and Paper I in this series.

\section{Numerical methods}\label{sec:num}
In this section we describe our numerical methods, simulation setup, and diagnostics.
\subsection{Equations and code}
As in Paper I, we present simulations run with a version of the code RAMSES \citep{tey02,tey06} that solves the equations of ideal MHD on a uniform cylindrical grid \citep[e.g.,][]{fau14},
\footnote{Freely available at \hyperlink{https://sourcesup.renater.fr/projects/dumses/}{https://sourcesup.renater.fr/projects/dumses/}
} under the cylindrical approximation \citep[no vertical gravity; see, e.g.,][]{arm98,haw01,sor12}. The continuity, momentum and induction equations solved by RAMSES are given by 
\begin{equation}
\label{eq:RAM1}
    \frac{\partial \rho }{\partial t}
    +\nabla\cdot(\rho {\bf u})= 0,
\end{equation}
\begin{equation}\label{eq:RAM2}
    \frac{\partial (\rho {\bf u})}{\partial t}
    +\nabla\cdot( 
        \rho {\bf u u}
        -{\bf B B}
    )
    +\nabla \left(
        P+\frac{\bf B\cdot B}{2}
    \right)
    =-\rho\nabla\Phi,
\end{equation}
\begin{equation}\label{eq:RAM3}
    \frac{\partial {\bf B}}{\partial t}
    +\nabla\cdot({\bf u B}-{\bf B u})
    =0,
\end{equation}
where $\rho,{\bf u},P,{\bf B}$ and $\Phi$ are the mass density, fluid velocity, gas pressure, magnetic field and gravitational potential, respectively. We supplement Equations \eqref{eq:RAM1}-\eqref{eq:RAM3} with an isothermal equation of state $P=c_s^2\rho$, for $c_s$ the purely constant isothermal sound speed. The sound speed serves as a direct proxy for temperature and disc thickness in our idealised model.

Once again we approximate relativistic effects by utilizing a Paczynski-Wiita gravitational potential. With the exclusion of vertical gravity, this potential is given in cylindrical polars $(r,\phi,z)$ by
\begin{equation}\label{PWpot}
    \Phi=\dfrac{-GM}{r-2r_g}.
\end{equation}
With the gravitational constant $G$, central black hole mass $M$ and speed of light $c$ set to one, the gravitational radius $r_g=GM/c^2$ and frequency $\omega_g=c^3/(GM)$ define code units for space and time. Unless otherwise stated, velocities are then given in units of $c$.

Particle orbits with angular velocity $\Omega_\text{PW}$ determined by the force balance $r\Omega_\text{PW}^2=\partial_r\Phi$ deviate from the simple Keplerian rotation law $\Omega_K\propto r^{-3/2}$. In particular, our choice of $\Phi$ obliges the square of the horizontal epicyclic frequency $\kappa^2=2\Omega[2\Omega +r\partial_r\Omega]$ to fall below zero in the inner disc, defining the innermost stable circular orbit (ISCO) at a radius $r_\text{ISCO}=6r_g$. This non-monotonic variation in turn produces the maximum in $\kappa$ that defines a trapping region for r-modes (see fig. 1 in Paper I). Finally, the orbital period at the ISCO is given by $T_\text{orb}\sim61.56\omega_g^{-1}$.

\subsection{Initial conditions}\label{sec:ICs}
We begin with simulations of circular discs, initializing pure rotation on cylinders with $u_\phi=r\Omega_\text{PW}.$ The saturated states established in these circular disc simulations provide points of comparison, as well as initial conditions for runs in which we impose non-circular streamlines.  
As in Paper I, we generate disc eccentricity solely with our choice of outer radial boundary condition (see Section \ref{sec:BC}). 

We place the inner boundary at a smaller radius than the ISCO, within the plunging region; the radial domain is $[r_0,r_1]=[4r_g,18r_g]$ in most of our runs. In all cases we include the full azimuthal range. Unless otherwise stated, we place our vertical boundaries at $z=\pm H$, where $H=c_s/\Omega_\text{PW}$ is the isothermal scale height at $8r_g$ (a radial location close to the expected r-mode trapping region). For the value $c_s=0.03c$ taken in all of our simulations, $H\sim0.5r_g$, implying an aspect ratio of $H/r\sim 0.064$ at $r=8r_g$. We have taken a larger sound speed than those used in Paper I (where $c_s=0.01-0.02c$), in order to allow greater angular resolution. Inertial wave trapping worsens with increasing $c_s$ (and hence disc thickness), but large azimuthal resolutions are necessary to resolve MRI turbulence \citep[e.g.,][]{sor12}. As noted in Paper I, such a thin disc may not provide an appropriate description for the SPL emission state, during which large accretion rates make the flow geometry uncertain \citep[see, e.g.,][]{lao89,esi97,Nay0}. In any case, a complete model will need to include dissipation in a hot corona, which is excluded by our focus on the mid-plane.

Initially, we set the density to a constant $\rho_0$ outside $r_\text{ISCO},$ and to a floor value within. In order to avoid depletion of the disc due to turbulent transport, we follow \cite{fau14} in adding a source term to the continuity equation. Outside $7r_g,$ this source term introduces mass at a rate 
\begin{equation}
    \frac{\partial \rho}{\partial t}
    =-\frac{(\rho-\rho_0)}{\tau},
\end{equation}
where $\tau$ is taken as ten times the local orbital timescale. 

With regard to magnetic fields, we set $B_r=B_\phi=0$ and initialize purely vertical fields with radial profiles similar to the zero-net-flux configurations considered by \cite{sor12}. Specifically, we set
\begin{equation}
    B_z=B_0S\Omega\sin[2\pi(r-r_\text{ISCO})],
\end{equation}
where $S=1$ if $6<r/r_g<16$ and $0$ otherwise, and $B_0
$ scales the magnetic field such that the wavelength of the fastest growing MRI mode ($\lambda_\text{MRI}=2\pi\sqrt{16/15}V_{\text{A},z}/\Omega$, where $V_{\text{A},z}=B_z/\sqrt{\rho}$) remains $\leq H$ throughout the simulation domain. This configuration produces a plasma beta (ratio of gas to magnetic pressure) with a minimum of $\beta\sim80$ close to the ISCO.

On top of this MRI-unstable background, we impose white-noise velocity perturbations $|\delta u_i|\leq 10^{-6}c_s$. We then evolve the MRI-turbulent, circular disc to a saturated, quasi-steady state before `turning on,' at $t=100T_\text{orb}$, an outer boundary condition that generates the eccentricity. We finally run both the forced and un-forced simulations for an additional $100T_\text{orb}$.

\subsection{Boundary conditions}\label{sec:BC}
We implement periodic azimuthal boundary conditions, and radial boundary conditions similar to those implemented in Paper I. A `diode' outflow condition at the inner boundary sets the radial mass flux in the ghost cells to the value in the innermost active cell if that value is negative (inflowing), or to zero otherwise. Meanwhile, we calculate the azimuthal velocity perturbation to the background orbital motion at the last active cell, and add this to an extrapolation of $r\Omega_\text{PW}$ in each ghost cell. Density, vertical mass flux and magnetic fields are simply matched to their values in the innermost cells of the active domain. We find that altering this inner radial boundary condition for the magnetic field has little effect on results.

At the outer radial boundary, we set $u_r=u_z=B_\phi=B_z=0$, calculate $B_r$ to satisfy the solenoidal condition, and match ${\bf u}=r\Omega_\text{PW}\hat{\boldsymbol{\phi}}$. In simulations of circular discs we simply set $\rho=\rho_0$ at $r_1$. On the other hand, to produce non-circular streamlines in a given simulation we impose a non-axisymmetric, precessing density profile, $\rho_E(\phi,t)$, in the outer radial ghost cells. As described in section 3.4 of Paper I, a non-axisymmetric density profile enforced in the ghost cells produces a pressure gradient in the outer disc, in turn forcing the inward propagation of eccentricity. The forcing profile $\rho_E(\phi,t)$ is chosen to be an eigenmode of the non-linear (Newtonian) secular theory, evaluated at the outer boundary.\footnote{
A non-linear theory is necessitated by the amplitude of forcing used in our simulations. Static sinusoidal profiles for density (associated with an $m=1$ linear mode) also induce eccentricity, but we have found that density profiles and precession frequencies from more precise non-linear calculations produce a much cleaner quasi-steady state.}

We calculate these eigenmodes (and their associated precession frequencies) semi-analytically, following the procedure described by \citet{bar16}, and assume a ratio of outer to inner boundaries $r_1/r_\text{ISCO}=3$. The fundamental non-linear solutions can be uniquely identified by the maximum eccentricity in the eigenmode, denoted by $A_f.$ However, we stress that the actual eccentricity produced in the simulation domain is significantly lower than $A_f$, first because we only force the non-axisymmetric density profile in the ghost cells, and the resulting perturbations in the active domain are smaller amplitude. Additionally, the simulations (unlike the semi-analytical secular theory) are relativistic, and permit continual propagation through the inner boundary. Lastly, damping by dissipation, and more importantly non-linear steepening and potential shocking of the eccentric wave near the ISCO reduce the eccentricity amplitude (see Section \ref{sec:ecc}).

We implement periodic vertical boundary conditions, which are appropriate for the cylindrical model. In Paper I we found that these lead to the formation of laminar `elevator flows' (steady columns of constant $u_z$) as r-modes' growth saturates. Enforcing a rigid lid at $z=\pm H$ halts the formation of these (presumed) numerical artifacts, and in hydrodynamic simulations has little effect on the dynamics other than to force the excited inertial oscillations to form as global standing modes (rather than vertically travelling waves). We find that fluctuations associated with MRI turbulence disrupt the elevator flows in all but the most strongly eccentric discs, and so the need for an artificially rigid vertical boundary is less compelling.

In the absence of vertical gravity, periodic vertical boundary conditions usually preclude the formation of coherent standing waves, which generally need to be inserted by hand (see Appendix A1 in Paper I). We do not fine-tune our initial conditions, and so do not recover standing mode oscillations. But our simulation setup is still sufficient to demonstrate the excitation of radially confined inertial waves. We note that periodic vertical boundary conditions are only formally appropriate under the cylindrical approximation, and can be problematic in stratified simulations \citep[e.g.,][]{sal16}. Given the apparent connection between HFQPOs and coronal plasma, future investigations including vertical structure should certainly consider using boundary conditions that allow the outflow of mass and magnetic flux.

\subsection{Diagnostics}\label{sec:diag}
We use several of the same diagnostics used in Paper I, first of all denoting volume, azimuthal and vertical averages for a given quantity $X$ by
\begin{align}
    \langle X\rangle_V    &= \dfrac{\int_V X dV}{\int_VdV},\\
    \langle X\rangle_\phi &= \dfrac{1}{2\pi}\int_0^{2\pi} X \text{d}\phi,\\
    \langle X\rangle_z    &= \dfrac{1}{2H}\int_{-H}^H X \text{d}z,
\end{align}
where the volume $V$ covers the radial range $[6r_g,16r_g]$. Angular momentum transport in MRI-turbulent discs is typically quantified by considering the Maxwell and Reynolds stresses, and so we define
\begin{align}
    \alpha_\text{R}&=\frac
    {\langle\rho u_r\delta u_\phi\rangle_V}{\langle P\rangle_V},
\\
    \alpha_\text{M}&=
    -\frac{\langle B_rB_\phi\rangle_V}{\langle P\rangle_V},
\end{align}
where $\delta u_\phi=u_\phi-r\Omega_\text{PW}.$ We similarly write the volume-averaged plasma beta parameter as
\begin{equation}
    \beta=\frac{\langle P\rangle_V}{\langle P_M\rangle_V},
\end{equation}
where $P_M=|{\bf B\cdot B}|/2$ is the magnetic pressure.

The quantity $\langle\rho u_r^2\rangle_V/\langle P\rangle_V$ provides a good quantitative diagnostic for the strength of disc eccentricity, since eccentric deformations dominate the radial kinetic energy  \citep{pap05b}. The complex eccentricity (defined as $E=e\exp[\text{i}\varpi],$ where $e$ and $\varpi$ are the eccentricity and longitude of pericentre) is another measure but is non-trivial to calculate for the twisted distortions in our simulations, because of the non-linearity of the flow and the presence of turbulent fluctuations. For simplicity, we instead use the estimate
\begin{equation}\label{eq:Eb}
    \tilde{E}=\frac{\int \langle u_r\rangle_z\cos\phi\text{d}\phi}
    {\pi\langle u_\phi\rangle_{\phi,z}},
\end{equation} 
calculated after shifting (globally) in $\phi$ to account for the retrograde precession of the distorted disc. This average tracks the imaginary part of $E$, but we stress that it is correct only to \emph{first order} in eccentricity, and loses accuracy very close to the ISCO where the eccentric waves steepen and orbits come close to intersecting.

In the hydrodynamic simulations presented in Paper I, we found an appropriate diagnostic for r-mode excitation in $\langle \rho u_z^2\rangle_V$, as well as the fraction of vertical kinetic energy contained within an annular domain encompassing the r-mode trapping region ($r\in[7r_g,9r_g]$). This diagnostic paints a murkier picture in MRI-turbulent simulations, because the turbulence itself contributes to the vertical kinetic energy. We therefore follow \cite{ReM09} in relying most heavily on timing analysis to search for oscillation excitation. Specifically, we consider the power spectral density (PSD), defined as $P(\omega)\propto|\mathcal{F}(f)|^2,$ where $\mathcal{F}(f)$ is the Fourier transform of the signal $f(t)$. $P(\omega)$ quantifies the power in a given oscillation frequency $\omega.$ We primarily take $f$ to be a suitable average of the radial mass flux, and use angular frequencies in units of $\omega_g$ throughout.

\subsection{Hydrodynamic test simulations}\label{sec:tst}
In most of the simulations described in this paper, we use boundary conditions that force eccentricity more strongly than in Paper I, where quite weak eccentric structures were sufficient to excite r-modes. The inward-propagating eccentric waves in our MHD simulations, while producing only modest eccentricities ($e\lesssim0.1$), are nonetheless sufficiently non-linear to steepen and circularise near the ISCO \citep{lyn19}. In addition to this natural, non-linear decrease in eccentricity relative to linear predictions, we expect the eccentric waves to suffer turbulent damping due to the MRI. An interesting question is which of the two effects dominates.

To determine the degree to which the distortions' propagation is affected by MHD turbulence vis-a-vis non-linear, conservative circularisation, we have run 2D, purely hydrodynamic simulations using the same sound speed, resolution and forcing boundary conditions as in 3D. These simulations are described in Appendix \ref{sec:2Dh} and also discussed in Section 3.2.1. In summary, the hydrodynamic 2D simulations exhibit nearly identical disc deformations to their MHD counterparts, suggesting that the turbulent damping of eccentricity does not play a significant role, and that hydrodynamic non-linear effects control the eccentric wave evolution near the ISCO.

\section{Simulation results}\label{sec:res}
In this section we present the results of our simulations. Table \ref{tab:trExM} lists the (i) simulation label, (ii) resolution, (iii) forcing amplitude for eccentricity, (iv) Maxwell stress, (v) Reynolds stress, (vi) plasma beta, (vii) volume-averaged radial kinetic energy density (normalised by the kinetic energy associated with sonic motion), and maxima in $\tilde{E}$ both throughout the entire domain (viii) and near the trapping region (ix). As described in Section \ref{sec:ICs}, we use the saturated states achieved by $t=100T_\text{orb}$ in the circular runs (zn, znHR, znL, zn2h) as the initial condition for simulations in which we force eccentricity from the outer boundary. We then run both the resulting non-circular discs, as well as the original circular discs, for an additional $100T_\text{orb}.$ The values listed in Table \ref{tab:trExM} have been averaged over the last $50T_\text{orb}$ of each simulation. 

The resolution used in our runs provides $16$ vertical grid cells per $H,$ and a grid cell aspect ratio of d$r$:$r$d$\phi$:d$z\sim$2:2:1 at $r=8r_g$. The run zn is our fiducial simulation of a circular disc, which we use to produce the initial condition for most of our eccentric disc simulations. In znHR we have doubled both radial and azimuthal resolution (due to numerical expense, this simulation has only been run for $100T_\text{orb})$. Meanwhile in znL we have extended the radial domain to $30r_g$, and in zn2h and zn2hAf37 we have doubled the vertical extent (to $z\in[-2H,2H]$). 

\begin{table*}
\centering
    \caption{Table listing (i) simulation label, (ii) resolution, 
    (iii) maximum eccentricity of the non-linear eigenmode used to generate the density profile enforced at the outer boundary, 
    (iv) Maxwell stress 
    $\alpha_M=-\langle B_rB_\phi\rangle_V/\langle P\rangle_V$,
    (v)  Reynolds stress 
    $\alpha_R=\langle \rho u_r\delta u_\phi\rangle_V/\langle P\rangle_V$, 
    (vi) plasma beta,
    (vii) volume-averaged radial kinetic energy density,
    (viii) maximum in $\tilde{E}$ (plotted in Fig. \mbox{\ref{fig:ecc}}),
    and (ix) maximum in $\tilde{E}$ between $7r_g$ and $9r_g$. All values have been averaged over the last $50T_\text{orb}$.}\label{tab:trExM}
    \begin{tabular}{lcccccccccr} 
        \hline
        Label & 
        $N_r\times N_\phi \times N_z$ &
        $A_f$ & 
        $\alpha_M$ & 
        $\alpha_R$ &
        $\beta$ &
        $\langle \rho u_r^2\rangle_V/\langle P\rangle_V$ & 
        $\max[\tilde{E}]$ & 
        $\max_D[\tilde{E}]$ \\
        \hline
        \hline
        zn     & $200\times800\times32$  & $0$ & 0.019 & 0.009 & 20.7 & 0.052 & -- & --\\
        znHR   & $400\times1600\times32$ & $0$ & 0.021 & 0.009 & 20.9 & 0.052 & -- & --\\
        znL    & $384\times800\times32$  & $0$ & 0.020 & 0.010 & 20.6 & 0.056 & -- & --\\
        zn2h   & $200\times800\times64$  & $0$ & 0.030 & 0.010 & 13.5 & 0.060 & -- & --\\
        \hline
        znAf10   & $200\times800\times32$  & $0.10$  & 0.019 & 0.009 & 20.8  & 0.058 & 0.012 & 0.007\\
        znAf30   & $200\times800\times32$  & $0.30$  & 0.017 & 0.012 & 22.9  & 0.161 & 0.045 & 0.024\\
        znAf35   & $200\times800\times32$  & $0.35$  & 0.012 & 0.016 & 29.6  & 0.250 & 0.058 & 0.028\\
        znAf37   & $200\times800\times32$  & $0.375$ & 0.006 & 0.019 & 53.7  & 0.315 & 0.068 & 0.030\\
        znAf40   & $200\times800\times32$  & $0.4$   & 0.002 & 0.025 & 133.9 & 0.421 & 0.077 & 0.033\\
        zn2hAf37 & $200\times800\times64$  & $0.375$ & 0.019 & 0.021 & 20.4  & 0.329 & 0.067 & 0.030\\
        \hline
     \end{tabular}
\end{table*}

\begin{figure*}
    \centering
    \includegraphics[width=.95\textwidth]{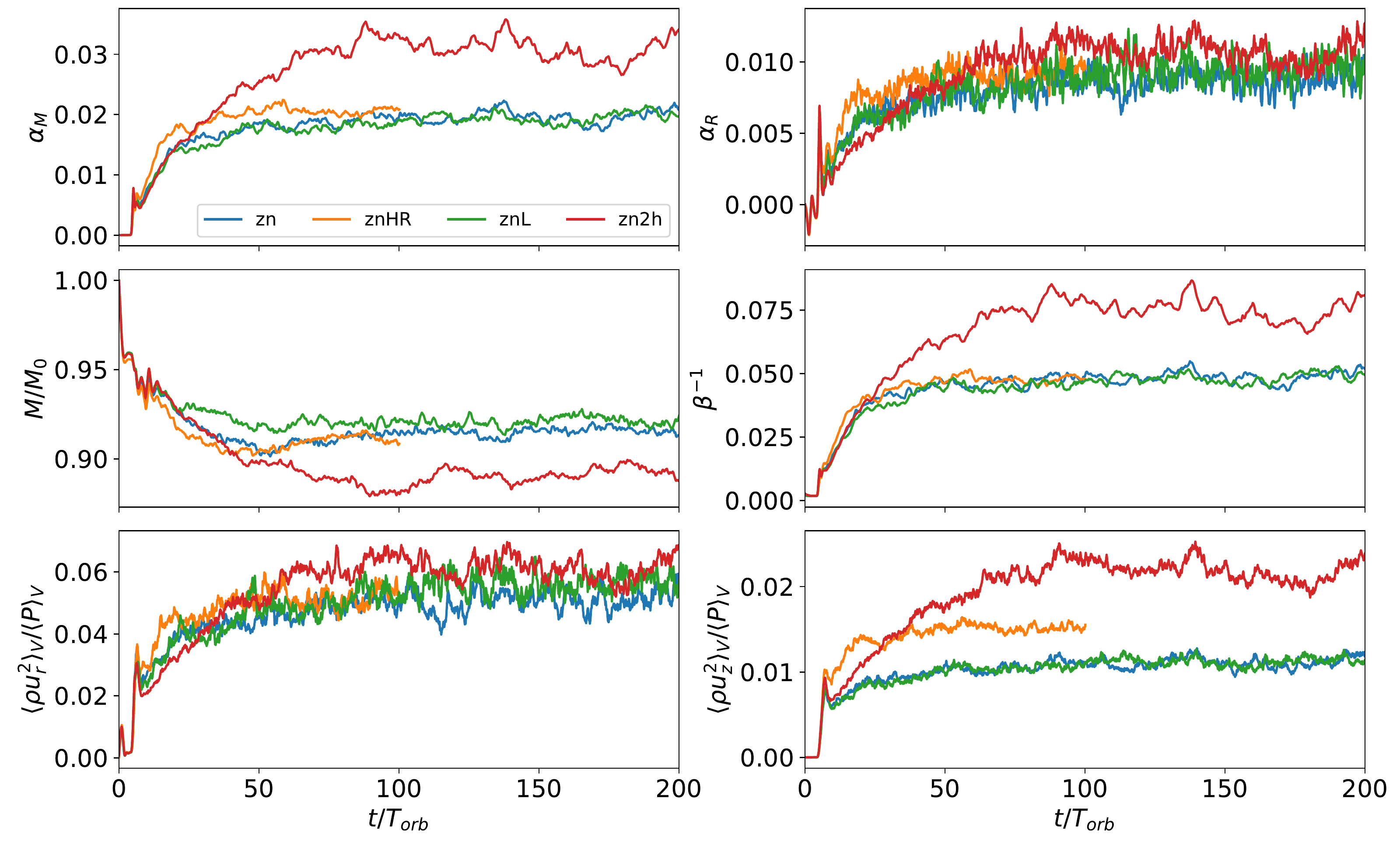}
    \caption{Volume-averaged quantities for our fiducial simulations of circular discs. Top left: Maxwell stress. Top right: Reynolds stress. Middle left: Mass fraction. Middle right: inverse plasma beta. Bottom left: radial kinetic energy. Bottom right: vertical kinetic energy. These history plots illustrate the saturated, quasi-steady states after $t=100T_\text{orb}$ that provide both the initial conditions and a point of comparison for our eccentric disc simulations.}\label{fig:circHist}
\end{figure*}

\subsection{Circular disc simulations}\label{sec:circ}
Although our primary focus is on eccentric discs, in this section we pause to describe the results of our circular disc simulations. These circular runs are important to consider, first because the quasi-steady states which they achieve provide the initial conditions for our eccentric disc simulations. Second, they provide an opportunity to check the performance of our code, which has been used to explore the linear \citep{lat15} and non-linear \citep{fau14} evolution of the global MRI in Newtonian, but not relativistic contexts. Lastly, our circular disc simulations provide an important point of comparison necessary for identifying r-modes in our timing analysis of the eccentric runs (Section \ref{sec:ecc}). 

\subsubsection{Volume-averaged quantities}
The plots in Fig. \ref{fig:circHist} show volume-averaged quantities over time for our simulations of circular discs. Fig. \ref{fig:circHist} (top left) illustrates the saturation of the Maxwell stress at values $\alpha_M\sim0.02-0.03$. These values are consistent with analogous simulations run by \cite{sor12} with a Newtonian potential. We note that we have also run simulations initialized with a net-flux vertical field, but, following an initial transient these simulations produce nearly identical saturated states to the zero-net-flux runs. We attribute this outcome to advection of the net-vertical flux through the ISCO and inner boundary, since in test simulations with a Newtonian potential we have reproduced the elevated Maxwell stresses observed by \cite{sor12} in their net-flux runs. Future simulations might consider the effects of a more sustainable source of poloidal magnetic field on disc variability.

\begin{figure*}
    \centering
    \includegraphics[width=\textwidth]{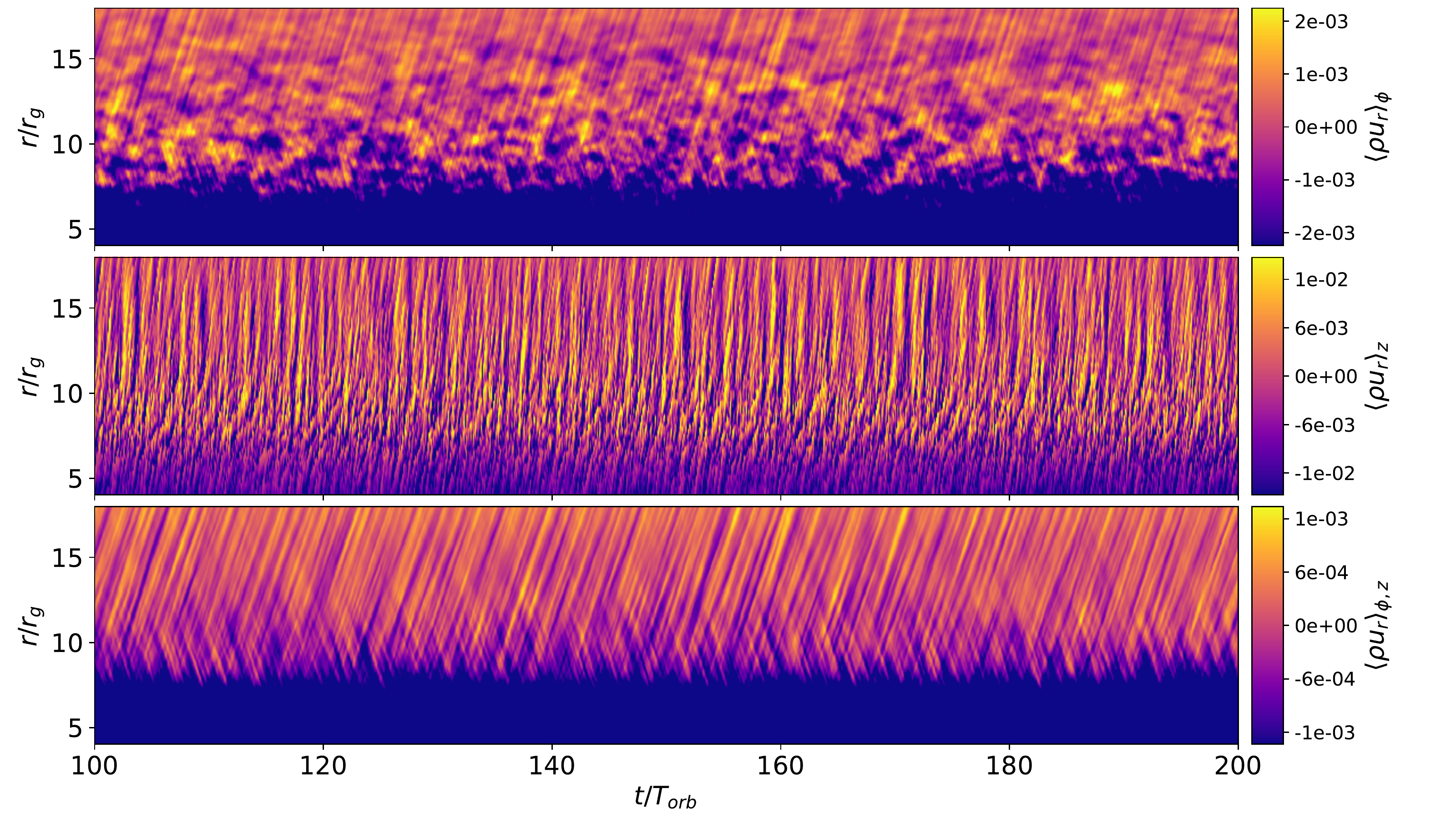}
    \caption{Spacetime diagrams illustrating azimuthally averaged (top), vertically averaged (middle) and both azimuthally and vertically averaged (bottom) radial mass flux over the last $100T_\text{orb}$ of runtime for simulation zn. These spacetime diagrams indicate periodic behavior over a range of radii and frequencies, but provide no evidence of trapped wave excitation.} \label{fig:zn_spct}
\end{figure*}

Fig. \ref{fig:circHist} (top right) shows somewhat larger Reynolds stresses than are typically observed in simulations of the MRI, indicating saturated ratios $\alpha_M/\alpha_R\sim 2-3$ (rather than the usual $4-6$). As described in Section \ref{sec:circTime}, the additional contribution to the Reynolds stress may be partially due to inertial-acoustic f-modes (i.e., density waves) originating from the inner disc. Meanwhile, Fig. \ref{fig:circHist} (middle left) shows an equilibration of the total mass (normalized by its initial value), balancing mass accretion through the inner boundary against mass input from the source term in the continuity equation. Fig. \ref{fig:circHist} (middle right) shows saturation of the inverse plasma beta at $\beta^{-1}\sim 0.05$, while the bottom panels in Fig. \ref{fig:circHist} show similar routes to saturation for the volume-averaged radial and vertical kinetic energy densities.

\begin{figure*}
    \centering
    \includegraphics[width=\textwidth]{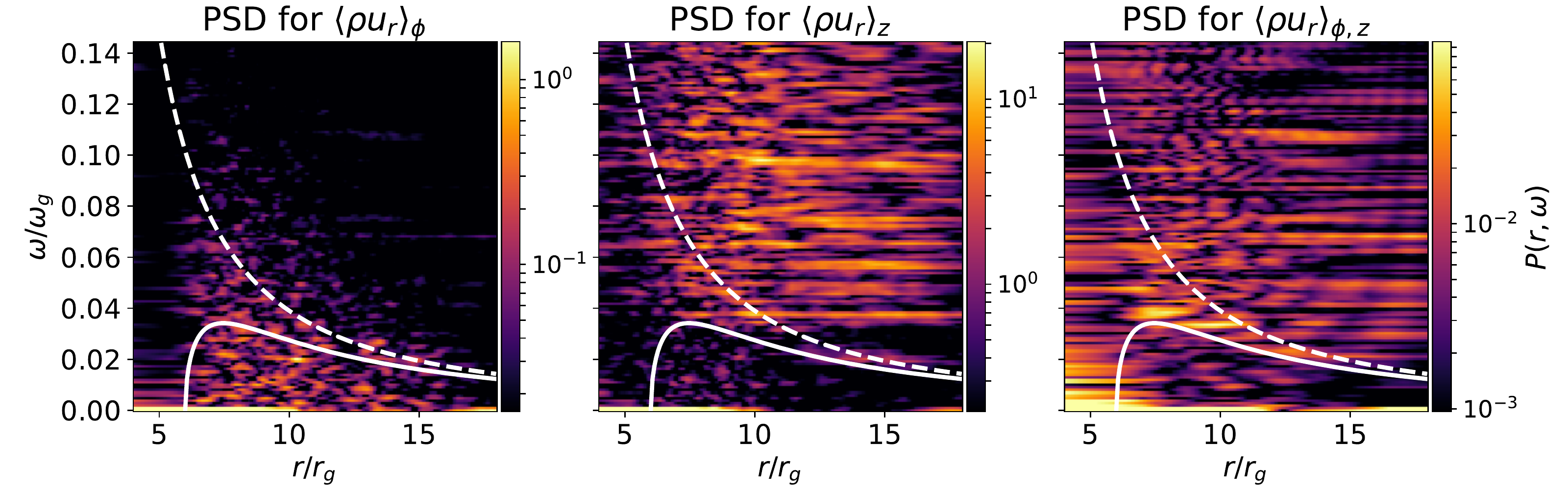}
    \caption{Power spectral densities calculated from the spacetime data shown in Fig. \ref{fig:zn_spct}. The solid and dashed white lines show the profiles of $\kappa_{PW}$ and $\Omega_{PW}$ associated with a Paczynski-Wiita potential. The left-hand PSD for $\langle \rho u_r\rangle_\phi$ reveals broad-band inertial power, the middle PSD for $\langle \rho u_r\rangle_z$ shows a spectrum of inertial-acoustic f-modes propagating at all radii, and the right-hand PSD for $\langle \rho u_r\rangle_{\phi,z}$ suggests the localization of an axisymmetric f-mode close to the maximum in $\kappa_{PW}$.} \label{fig:zn_pds}
\end{figure*}

Our circular disc simulations produce magnetic tilt angles \citep{Gua09} of $\theta_B=\arcsin(\alpha_M\beta)/2\approx11-12^\circ$, suggesting marginally resolved turbulence \citep{sor12}. As indicated by the values in Table \ref{tab:trExM} and Fig. \ref{fig:circHist}, znHR and znL (our simulations with larger resolution and radial domain, resp.) exhibit very similar saturated states to zn, although in znHR the vertical kinetic energy saturates at a marginally higher value. In zn2h, doubling the vertical extent of the disc produces larger stresses, matched by a decrease in $\beta$. This is unsurprising in light of unstratified shearing box simulations that, with the exclusion of vertical gravity, show a dependence of the turbulent stress on the vertical extent of the 
domain \citep{shi16}. The cylindrical approximation is invalid on scales larger than a scale height, however, and including a very large vertical extent would eliminate the unstratified model's advantage as an inexpensive numerical framework. In any case, our circular disc simulations fulfill our primary requirement of producing quasi-steady states characterized by sustained, generic MRI turbulence. Further, we observe very similar dynamics in zn2hAf37 (an eccentric disc simulation initialized from zn2h) and znAf37, its counterpart with the vertical range $z\in[-H,H]$.

\subsubsection{Timing analysis for zn}\label{sec:circTime}
In order to gain a qualitative understanding of the oscillations (or lack thereof) in the circular disc simulation zn, we consider spacetime diagrams in time and radius from the last $100T_\text{orb}$. Those shown in Fig. \ref{fig:zn_spct} illustrate radial profiles of radial mass flux. We consider mass flux rather than velocity in order to exclude fluctuations in the low-density region within the ISCO. To isolate modes with different character, we provide spacetime diagrams showing mid-plane profiles of $\langle \rho u_r\rangle_\phi$ (top), $\phi=0$ profiles of $\langle \rho u_r\rangle_z$ (middle) and profiles of $\langle \rho u_r\rangle_{\phi,z}$ (bottom). These averages work as dynamical filters, allowing us to home in on the oscillations of interest. Specifically, taking an azimuthal average isolates axisymmetric modes with azimuthal wavenumber $m=0$, which are of the most interest to the r-mode model for HFQPOs. Taking a vertical average, on the other hand, shifts focus to inertial-acoustic f-modes, since they have no vertical structure (i.e., vertical wavenumber $k_z=0$). Finally, $\phi-z$ averages isolate `fundamental' axisymmetric inertial-acoustic waves with $k_z=m=0$. 

We gain more quantitative insight from power spectral densities $P(r,\omega)$ computed from the spacetime data in Fig. \ref{fig:zn_spct}, which we show via the heatmaps in Fig. \ref{fig:zn_pds}. We normalize $P(r,\omega)$ arbitrarily, but saturate the colour-plots according to the largest peak at frequencies $\omega>0.005\omega_g$. The solid and dashed lines plot radial profiles for $\kappa_\text{PW}$ and $\Omega_\text{PW}$ calculated for particle orbits in a Paczynski-Wiita potential. These can differ from characteristic orbital and epicyclic frequencies calculated from the actual fluid flow, due to background pressure gradients, magnetic stresses, or a dynamic redistribution of angular momentum by oscillations in the disc (see Paper I). We find, however, that in our MHD-turbulent simulations, restoring forces from magnetic tension do more to modify inertial waves' frequencies and trapping than higher order modifications to $\kappa.$

Like \cite{ReM09}, we do not observe \emph{explicit} signatures of global trapped inertial waves in our simulations of circular discs. Qualitatively, the spacetime diagram for $\langle \rho u_r\rangle_\phi$ (top panel) indicates some variability in the inner disc, but no dominant axisymmetric oscillations. The broad-band power concentrated beneath the profile for $\kappa$ (at all radii) in the PSD shown in Fig. \ref{fig:zn_pds} (left) suggests that the fluctuations apparent in Fig. \ref{fig:zn_spct} (top) are inertial in nature, since local analyses predict that axisymmetric inertial waves are evanescent except where $\omega^2<\kappa^2$. The fact that this power extends to frequencies greater than $\kappa$ can be explained by an enhancement of small-scale inertial wave frequencies by Alfv\'enic restoring forces \citep[see, e.g.,][]{FL09,par18,dew18}. Figs. \ref{fig:zn_spct} (top) and \ref{fig:zn_pds} (left) do not exclude the presence of trapped r-modes, but the plots support the conclusions of \cite{ReM09} that the MRI does not preferentially excite them over the spectrum of small-scale inertial or, rather, `Alfv\'enic-epicyclic' waves. 

Meanwhile, the spacetime diagrams in Fig. \ref{fig:zn_spct} for $\langle\rho u_r\rangle_z$ (middle) and $\langle\rho u_r\rangle_{\phi,z}$ (bottom) illustrate the propagation of many inertial-acoustic waves. As described in Paper I, both axisymmetric and non-axisymmetric f-modes may be excited in the inner regions of black hole accretion discs. They are prone to viscous overstability, for instance, in viscous models of relativistic discs possessing a transonic inflow \citep[e.g.,][]{chan09,mir15}. In discs with a reflecting inner boundary, on the other hand, non-axisymmetric f-modes are subject to excitation via a transmission of wave energy at the corotation radius (where the pattern speed matches the orbital angular velocity), due to the profile of vortensity in a relativistic disc \citep{lai09,FL11,FL13}. 

The PSD for $\langle\rho u_r\rangle_z$ shown in Fig. \ref{fig:zn_pds} (middle) reveals a spectrum of f-modes across a broad range of $m$, this time visible as horizontal streaks. The PSD for $\langle\rho u_r\rangle_{\phi,z}$ (Fig. \ref{fig:zn_pds}, right) indicates that one of these can be identified as an axisymmetric f-mode with $\omega\gtrsim\max[\kappa],$ located at $r\sim 8r_g$ (close to where $\kappa$ achieves this maximum). A detailed analysis of the non-axisymmetric f-modes indicated by Fig. \ref{fig:zn_pds} (middle) is beyond the scope of this work, but PSDs calculated by taking spatial Fourier transforms in azimuth (rather than time) show that waves with azimuthal wavenumber $m=2-4$ dominate. This is not inconsistent with the excitation of non-axisymmetric f-modes by the corotational instability considered by \cite{lai09}, which produces the largest growth rates for $m=2,3$. But in Paper I we found \citep[as did ][]{mir15} that placing the inner boundary within the ISCO and imposing an outflow boundary condition greatly reduces the impact of this instability, which relies on reflection at the inner boundary. Further, \cite{FL11} found that while corotationally excited f-modes' propagation remains relatively unaltered by large-scale magnetic fields, their growth rates may be reduced. These considerations suggest that the MRI turbulence itself is responsible for exciting the inertial-acoustic waves \citep[see, e.g.,][]{Hei09a,Hei09b,Hei12}.

\subsection{Eccentric disc simulations}\label{sec:ecc}
In this section we present the main results of the paper, describing our eccentric disc simulations. As detailed in Sections \ref{sec:ICs} and \ref{sec:BC}, we initialize these runs with the quasi-steady states achieved by zn and zn2h at $100T_\text{orb}$, and then turn on the eccentricity-forcing outer radial boundary condition. We discuss the inward propagation of eccentricity caused by this boundary condition in Section \ref{sec:eprop}, and its dynamical consequences in Section \ref{sec:dcon}.

\subsubsection{Eccentricity propagation}\label{sec:eprop}
In each forced simulation, the precessing, non-axisymmetric density profile enforced at the outer radial boundary induces a travelling eccentric wave that propagates inward at roughly $c_s/2$. Reaching the ISCO within $\sim10-20T_\text{orb}$, these waves saturate as steady, slowly precessing distortions. Because the eccentric waves continually propagate through the ISCO and out of the simulation domain (rather than reflecting off an inner boundary), the disc deformations in all simulations manifest as twisted eccentric modes with a radially varying phase. 

\begin{figure}
    \centering
    \includegraphics[width=\columnwidth]{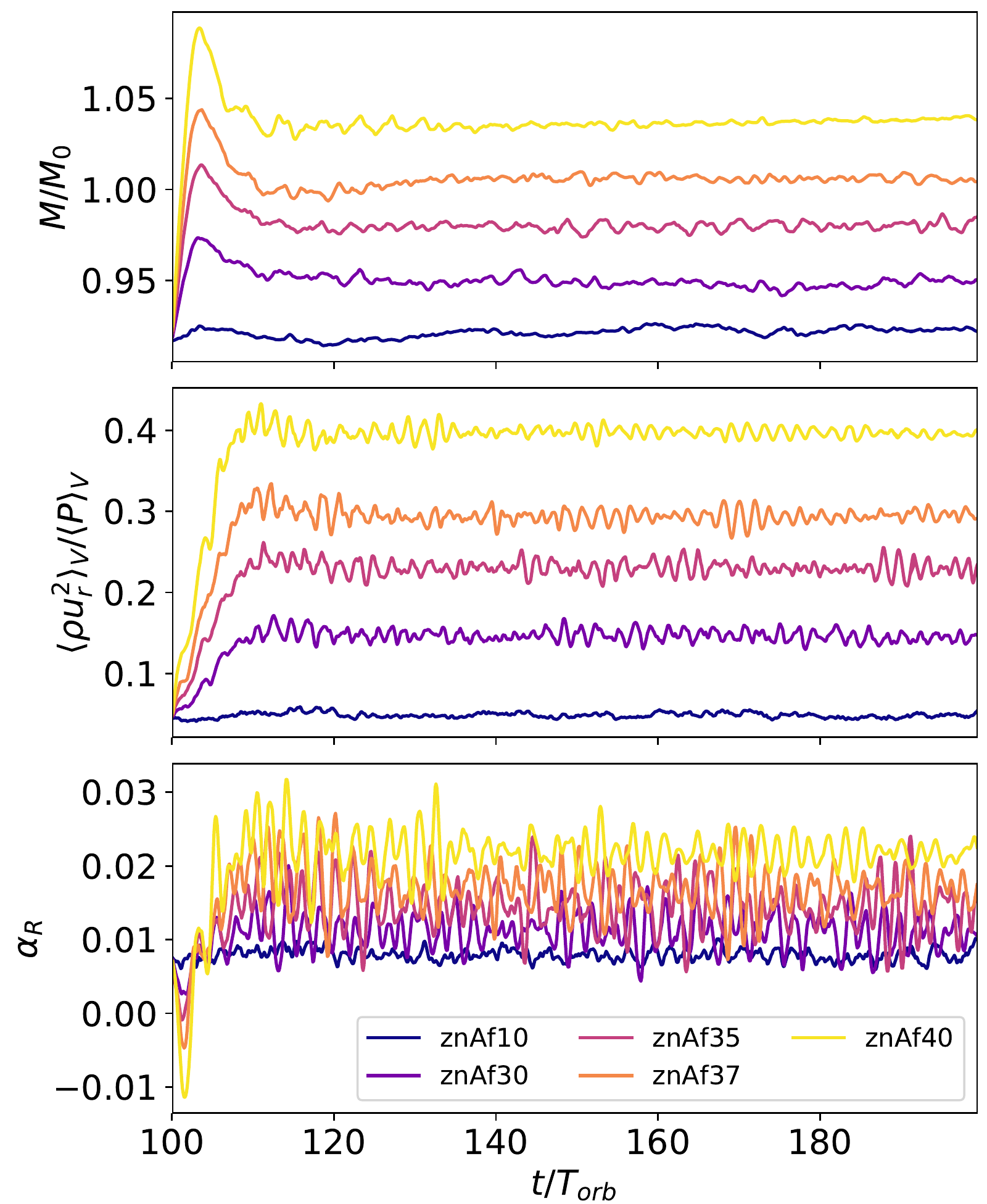}
    \caption{History plots showing mass fraction (top), volume-averaged radial kinetic energy density (middle), and Reynolds stress (bottom) in simulations znAf10-40 (in which we force eccentric distortion). The change and rapid saturation in these quantities from the initial values provided by the circular run zn illustrate the transition to quasi-steady, twisted eccentric discs, with larger distortions produced by larger forcing amplitudes.}\label{fig:MrKEaRhist}
\end{figure}

The volume-averaged quantities plotted in Fig. \ref{fig:MrKEaRhist} illustrate the rapid saturation of the eccentric distortions in simulations znAf10, znAf30, znAf35, znAf37 and znAf40. The top panel shows total disc mass (normalized by the total mass at $t=0$ initialized in zn), which increases marginally, due to the eccentricity-forcing outer boundary condition, before reaching a new equilibrium. Fig. \ref{fig:MrKEaRhist} (middle) plots $\langle\rho u_r^2\rangle_V/\langle P\rangle_V$ (a quantity dominated by the eccentric distortion), and similarly indicates a rapid transition to quasi-steady states with larger forcing amplitudes giving rise to larger eccentricities. As shown in the bottom panel, the twisted, travelling eccentric waves enhance the Reynolds stress $\alpha_R$, because they involve a net angular momentum flux.

\begin{figure}
    \centering
    \includegraphics[width=\columnwidth]{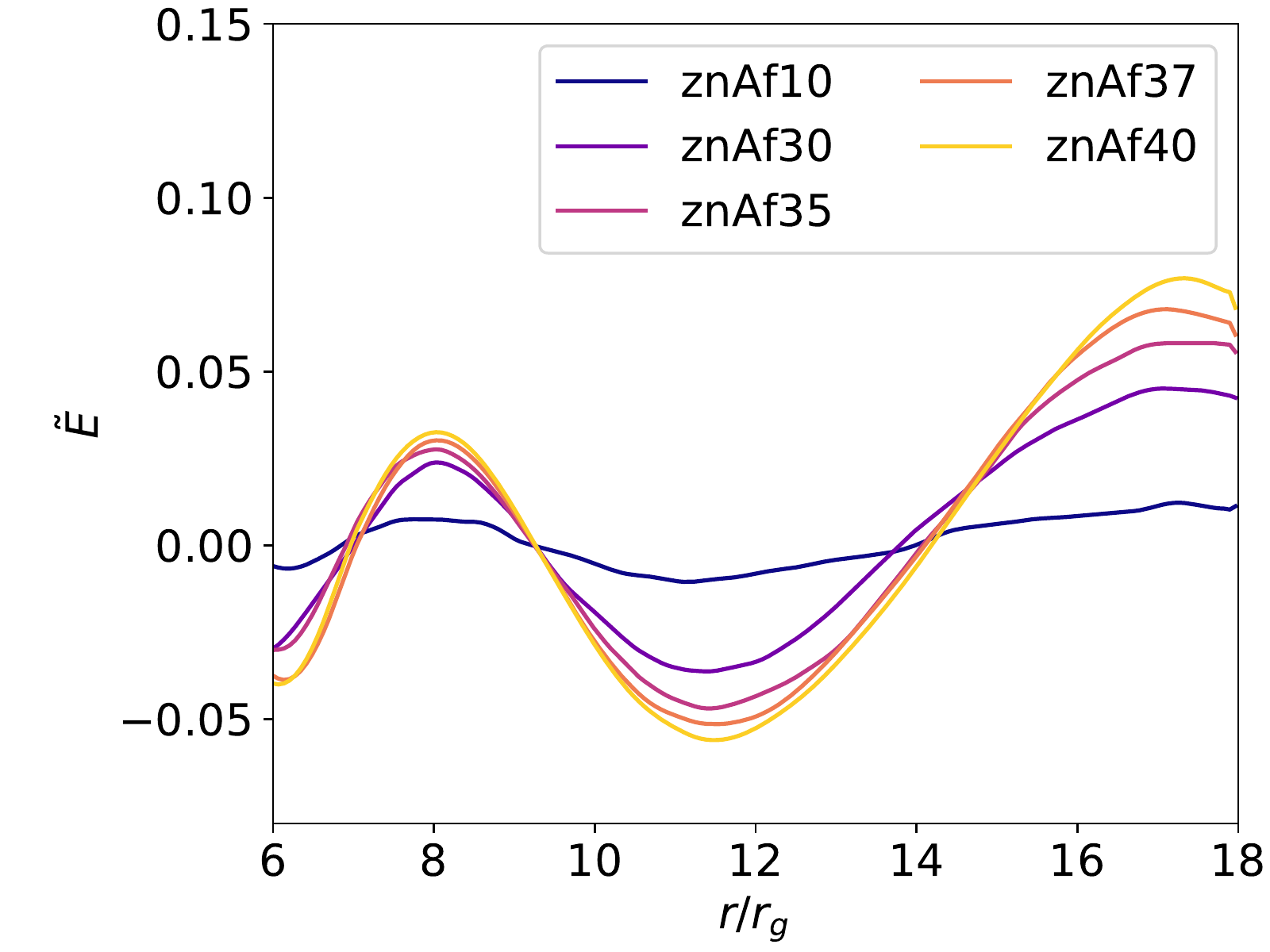}
    \caption{First-order estimate (see Equation  \ref{eq:Eb}) of the imaginary part of the complex eccentricity, $\tilde{E}\sim-\text{Im}[E]=-e\sin\varpi$, averaged over $50T_\text{orb}$.  Larger forcing amplitudes $A_f$ at the outer boundary produce larger eccentricities in the simulation domain. This linear approximation illustrates the basic radial structure of the eccentric distortions in our simulations, including a decrease in eccentricity near the ISCO due to non-linear wave steepening. We note that as a linear approximation, $\tilde{E}$ does not accurately capture this non-linear steepening of the eccentric waves, which is clearer in Fig. \ref{fig:streamax} (top).}
    \label{fig:ecc}
\end{figure}

Fig. \ref{fig:ecc} provides a clearer picture of the structure of the eccentric distortions in simulations znAf10-40, plotting radial profiles of $\tilde{E}$, a linear approximation to the complex eccentricity (see Equation \ref{eq:Eb}). We calculate these profiles at a given time $t$ after shifting globally in azimuth by $\delta\phi=|\omega_P|t,$ where $\omega_P<0$ is the precession frequency enforced at $r_1$. This produces profiles of $\tilde{E}$ that are in phase despite the different precession frequencies associated with different forcing amplitudes for eccentricity. The profiles in Fig. \ref{fig:ecc} have further been averaged over the final $50T_\text{orb}$ of each simulation.

Notably, the plot shows that the eccentric distortions decrease in amplitude toward the inner disc. This decrease disagrees with linear theory, which predicts an increase of eccentricity near the ISCO \citep[see ][or the more weakly forced simulations in paper I]{fer09}. However, a recent nonlinear analysis \citep{lyn19} shows that the steepening of the wave and the occurrence of near intersections between neighbouring streamlines may cause the eccentricity to decrease inwards, even if the angular-momentum flux is conserved. Furthermore, steepening of the wave can greatly enhance dissipation, causing an attenuation of the angular-momentum flux and therefore the wave amplitude.

One might also ask whether turbulent fluctuations in our simulations are also damping the eccentric waves' inward propagation. Indeed, \cite{fer09} found that turbulent damping, parametrised by a bulk viscosity, can hinder the propagation of warps and eccentricities. However, we have run 2D, purely hydrodynamic simulations with the same resolution, boundary conditions and forcing amplitudes as znAf10-40. These hydrodynamic runs show very similar distortions to their MHD counterparts (see Appendix \ref{sec:2Dh}), demonstrating that the decrease in eccentricity near the ISCO is due primarily to non-linear steepening of the eccentric waves, rather than to turbulent damping.

\begin{figure*}
    \centering
    \includegraphics[width=\textwidth]{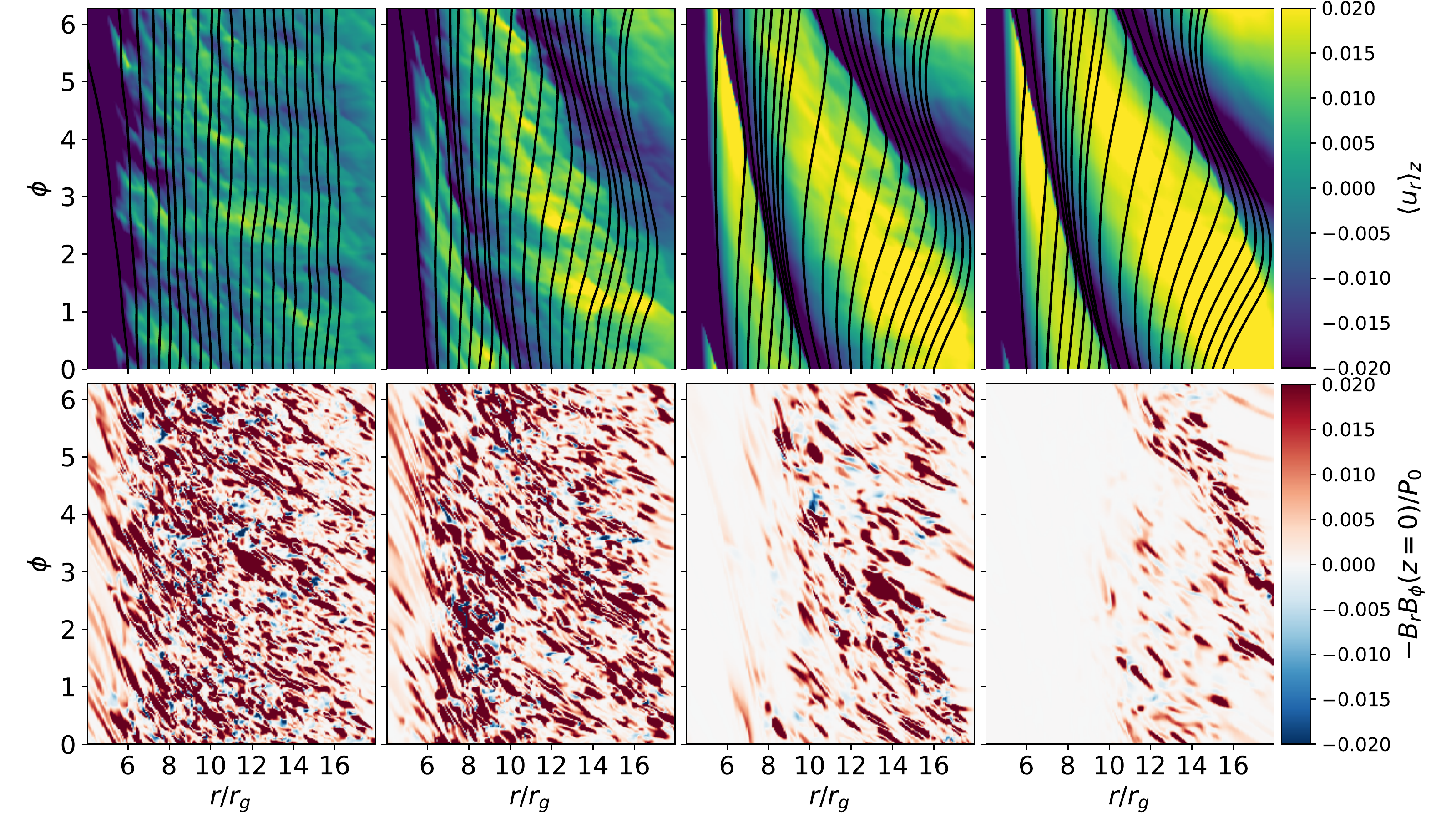}
    \caption{Unfolded polar snapshots showing vertically averaged radial velocity and overplotted streamlines (top), and mid-plane Maxwell stress normalized by $P_0=c_s^2\rho_0$ (bottom) for simulations (from left to right) znAf10, znAf30, znAf37, and znAf40. The snapshots are taken from simulations with larger and larger forcing amplitudes for eccentricity (from left to right), $50T_\text{orb}$ after turning on the outer boundary condition that drives the deformation. The top plots illustrate the eccentric distortion of the disc, while the bottom plots suggest a weakening of the MRI turbulence with increasing eccentricity.} \label{fig:streamax}
\end{figure*}   

The colour-plots in Fig. \ref{fig:streamax} (top) illustrate this steepening, showing snapshots of $\langle u_r\rangle_z$ (overlaid by streamlines calculated from the vertically averaged flow) taken $50T_\text{orb}$ after initializing our eccentric forcing in simulations (from left to right) znAf10, znAf30, znAf37 and znAf40. The colour-plots on the bottom of Fig. \ref{fig:streamax} show mid-plane snapshots of Maxwell stress taken at the same time. The streamlines depict the distortion caused by our outer boundary condition, since circular orbits appear as vertical lines on unfolded polar plots (although we note that equivalent radial deviations at larger radii imply smaller eccentricities). In the snapshots from the more strongly forced simulations, sharp changes in $u_r$ and very closely spaced streamlines illustrate the inherently non-linear wave steepening not captured by the linear approximation $\tilde{E}$ plotted in Fig. \ref{fig:ecc}.

\subsubsection{Dynamical consequences of eccentricity}\label{sec:dcon}
Broadly, we identify three dynamical regimes for the simulations with forced eccentricity, depending on the degree of distortion. For forcing amplitudes $A_f\lesssim0.35$ (which produce eccentricities of $e\sim 0.007-0.03$ near the trapping region), we do not observe clear signatures of trapped r-modes. However, eccentricities of $e\gtrsim0.03$ produced by larger forcing amplitudes do appear to excite trapped inertial waves, which leave distinct timing features in the PSD. Additionally, in this regime the eccentricity begins to significantly interfere with the MRI, reducing the Maxwell stress (as illustrated by the sequence of snapshots in Fig. \ref{fig:streamax}, bottom). In the third regime, even larger forcing amplitudes ($A_f\gtrsim0.4$) produce eccentric distortions that suppress the MHD turbulence within $r\lesssim10r_g$. Our simulations in this regime yield similar results to those obtained in hydrodynamic runs utilizing periodic vertical boundary conditions (see Paper I). In the following subsections, we discuss each of the regimes of behavior depicted in Fig. \ref{fig:streamax} in turn. 

\begin{figure}
    \centering
    \includegraphics[width=\columnwidth]{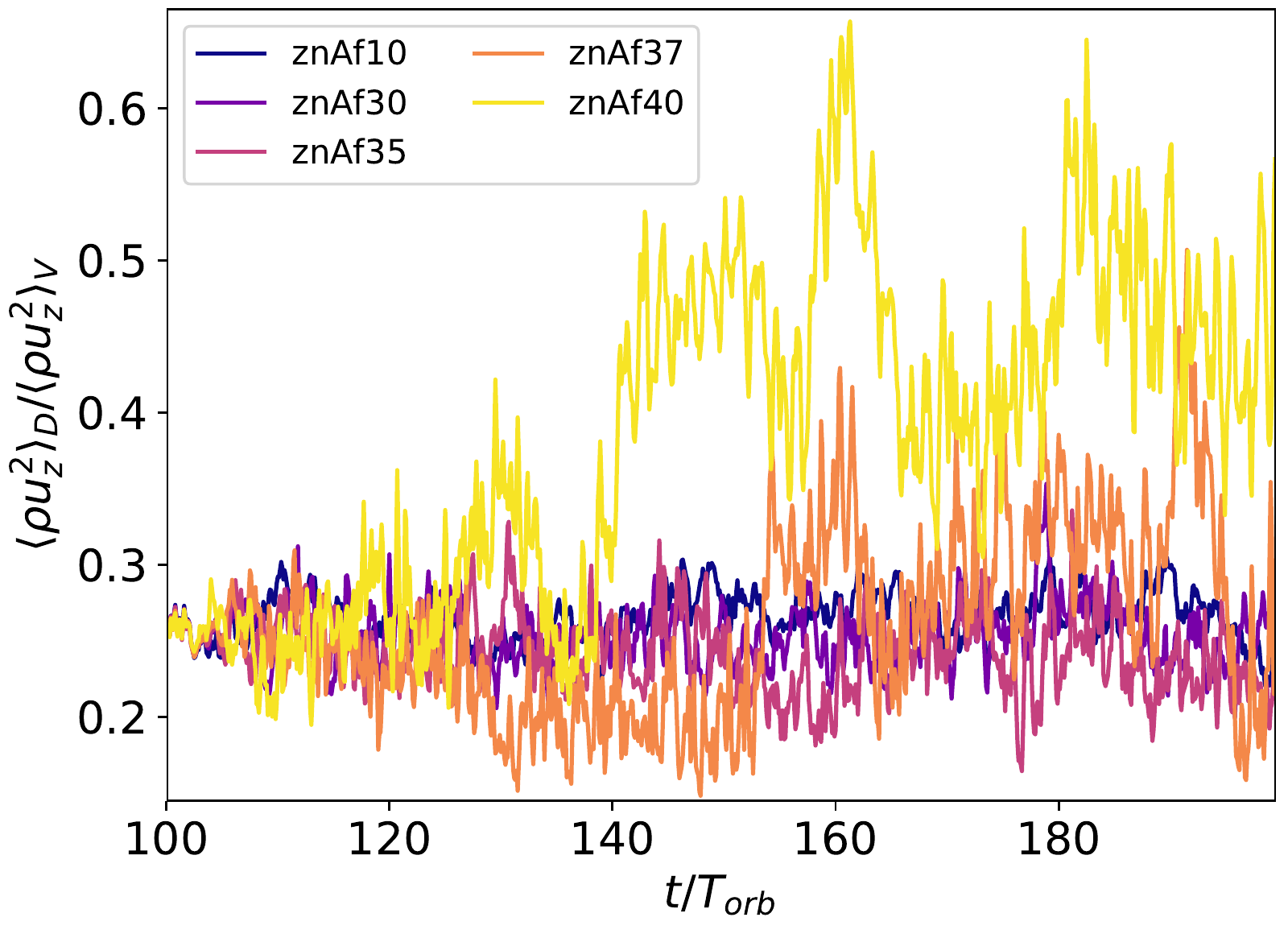}
    \caption{The fraction of vertical kinetic energy contained within the annular domain $D$ defined by $r\in[7r_g,9r_g]$. Simulations znAf10, znAf30, and znAf35 show little change from the saturated value taken from the circular run zn. On the other hand, znAf37 and znAf40 show larger deviations, hinting at enhanced variability due to trapped wave excitation.}
    \label{fig:vKEhist}
\end{figure}

\begin{figure*}
    \centering
    \includegraphics[width=\textwidth]{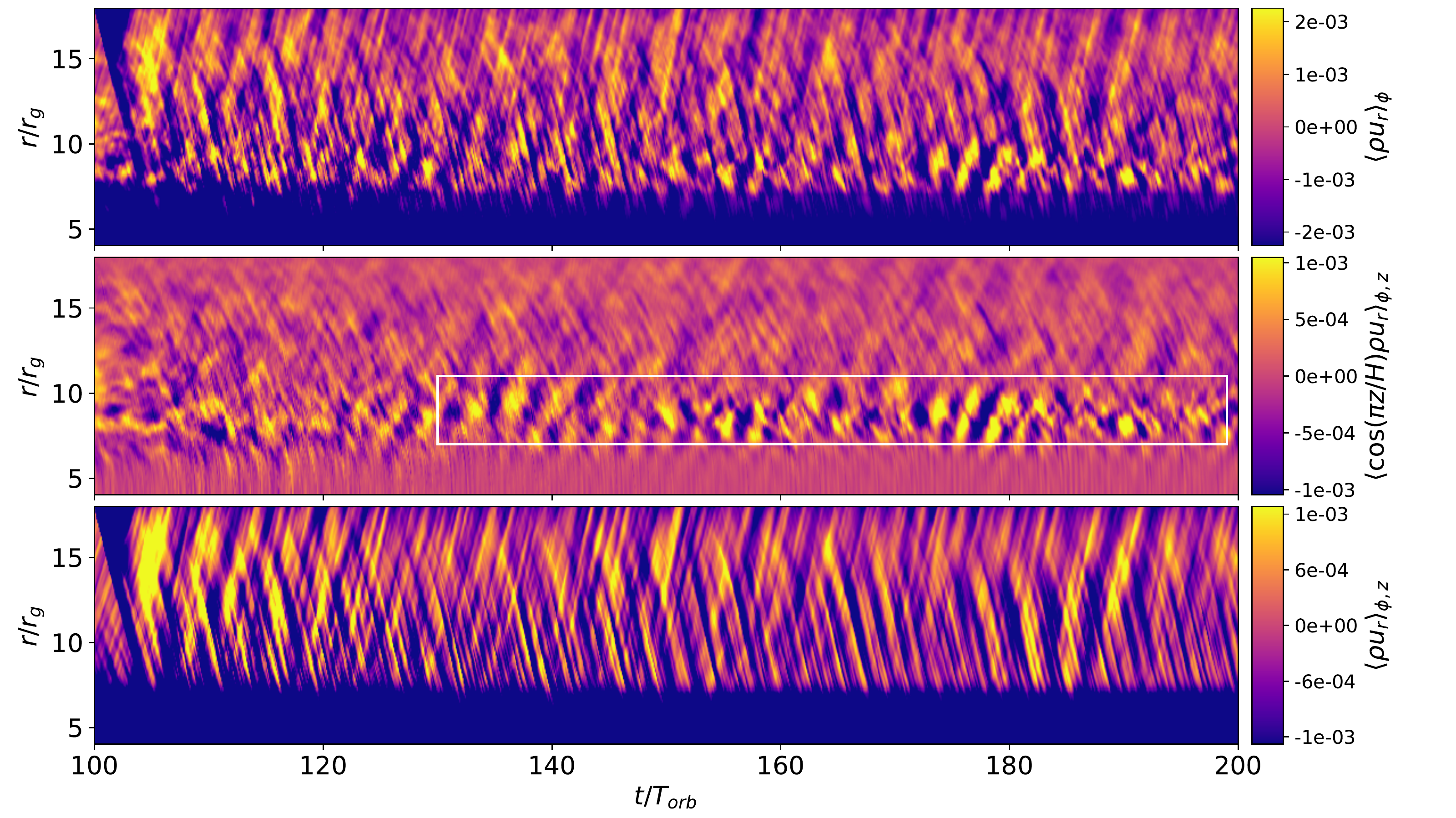}
    \caption{Spacetime diagrams showing radial profiles of $\langle \rho u_r\rangle_\phi,$ (top) $\langle \cos(\pi z/H)\rho u_r\rangle_{\phi,z}$ (middle) and $\langle \rho u_r\rangle_{\phi,z}$ (bottom) over time in simulation znAf37. The white box in the middle plot highlights variability not apparent in the circular disc simulation zn, which we associate with (magnetically enhanced) trapped inertial waves excited by the eccentricity in the disc.} \label{fig:znAf375_spct}
\end{figure*}

\vspace{1em}
{\flushleft\textbf{Regime I: low-eccentricity}}
\vspace{0.5em}

We first discuss simulations znAf10, znAf30 and znAf35, in which forcing boundary conditions induce eccentricities of $e\lesssim0.03$ near the expected r-mode trapping region. As illustrated by Fig. \ref{fig:streamax} (top left), the velocity perturbations associated with the eccentric distortion in znAf10 possess amplitudes comparable to those of the turbulent fluctuations. Consequently, the eccentric distortion appears only with a time-average (see the darkest curve in Fig. \ref{fig:ecc}). While the distortions in znAf30 and znAf35 are more visible, none of the three runs exhibit signatures of trapped inertial wave excitation.

This absence is indicated both by a lack of distinct variability in timing analyses like that shown in Fig. \ref{fig:zn_pds}, and by Fig. \ref{fig:vKEhist}, which plots the fraction of vertical kinetic energy contained within the annular domain $r\in[7r_g,9r_g]$. In Paper I, this fraction increased rapidly in simulations initialized with non-negligible eccentricity, as r-modes trapped near $\sim8r_g$ grew large enough in amplitude to dominate the total vertical kinetic energy. In contrast, znAf10, znAf30 and znAf35 show little to no change from the saturated value of $\sim0.25$ produced by zn. Averaged over the final $50T_\text{orb},$ the three simulations respectively produce kinetic energy fractions of $\sim0.27,0.25,$ and $0.24$.

Given that weaker or comparable eccentricities readily excited r-modes in the hydrodynamic simulations in Paper I, this absence indicates that turbulence driven by the MRI damps the oscillations. It is possible that trapped inertial waves excited by the eccentric structures in these simulations simply do not reach large enough amplitudes to stand out above the noise, but unlikely that such oscillations would coexist with turbulent fluctuations without some interference. Turbulent damping of r-modes would align with the understanding that most contributions from dissipation due to turbulence in an accretion disc will go toward damping oscillations \citep{lat06}.

\vspace{1em}
{\flushleft\textbf{Regime II: trapped wave excitation}}
\vspace{0.5em}

We now turn to the intermediate regime in which the eccentric distortions excite trapped waves but do not completely suppress MRI turbulence. Simulation znAf37 provides the fiducial example, with a curve in Fig. \ref{fig:vKEhist} that shows (highly variable) enhancements in vertical kinetic energy within the trapping region $r\in[7r_g,9r_g]$.

\begin{figure*}
    \centering
    \includegraphics[width=\textwidth]{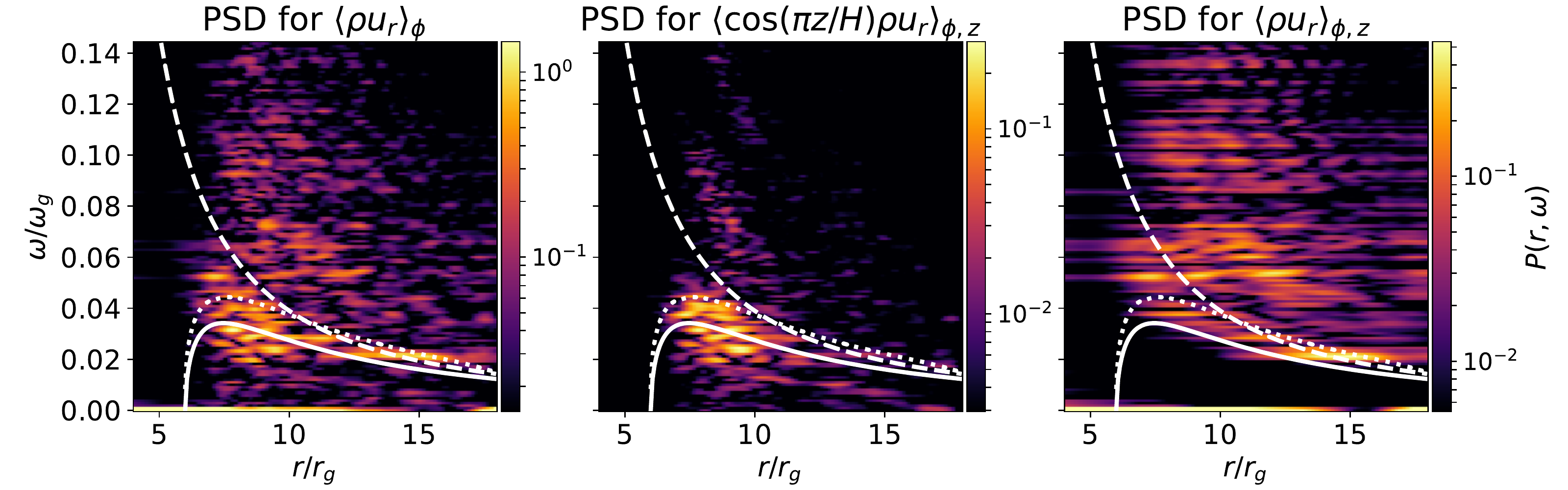}
    \caption{Power spectral densities calculated from the spacetime data shown in Fig. \ref{fig:znAf375_spct}, over the interval $t=120-200T_\text{orb}$. As in Fig. \ref{fig:zn_pds}, the solid and dashed white lines show the profiles of $\kappa_{PW}$ and $\Omega_{PW}$ associated with the Paczynski-Wiita potential, while the dotted line shows $\kappa_{PW}+\omega_{Az}$ for a mode with $k_z=\pi/H$. The peaks in power near $\max[\kappa]$ apparent in the PSDs for $\langle \rho u_r\rangle_{\phi}$ (left) and for
    $\langle \cos(\pi z/H)\rho u_r\rangle_{\phi,z}$ (middle) correspond to the trapped oscillations highlighted in the box in Fig. \ref{fig:znAf375_spct} (middle). The PSD calculated from spacetime data for $\langle \rho u_r\rangle_{\phi,z}$ (right) shows a more complicated spectrum of inertial-acoustic waves than was excited in the hydrodynamic simulations of Paper I.} \label{fig:znAf375_pds}
\end{figure*}

Like Fig. \ref{fig:zn_spct}, Fig. \ref{fig:znAf375_spct} shows spacetime diagrams calculated from different averages of the radial mass flux, this time from simulation znAf37. Since the slow precession of the eccentric disc dominates the vertical average $\langle\rho u_r\rangle_z$, we focus on radial profiles of $\langle \rho u_r\rangle_\phi,$ (top), $\langle \cos(\pi z/H)\rho u_r\rangle_{\phi,z}$ (middle), and $\langle \rho u_r\rangle_{\phi,z}$ (bottom). Azimuthal averages filter out the disc eccentricity, while the factor of $\cos(\pi z/H)$ removes vertically homogeneous features (i.e., f-modes). This factor also has the disadvantage of removing inertial waves with higher vertical wavenumbers, but still serves to isolate the axisymmetric, vertically structured oscillations of interest. Fig. \mbox{\ref{fig:znAf375_pds}} shows PSDs calculated from the spacetime data pictured in Fig. \mbox{\ref{fig:znAf375_spct}}, sampled between $t=120-200T_\text{orb}$.

The spacetime diagrams in Fig. \ref{fig:znAf375_spct} show enhanced variability over those in Fig. \ref{fig:zn_spct}. Specifically, the oscillations at $\sim 8r_g$ circumscribed by a white box in Fig. \ref{fig:znAf375_spct} (middle) produce features near the maximum of the epicyclic frequency in the PSDs shown in Fig. \ref{fig:znAf375_pds} (left and middle). In addition to $\kappa_{PW}$ (white line), the plots in Fig. \ref{fig:znAf375_pds} include radial profiles of $\kappa_{PW}+\omega_{Az},$ where $\omega_{Az}=\pi \tilde{V}_{Az}/H$ is an estimate of the Alfv\'en frequency for modes with $k_z=\pi/H$, calculated using a time-average of
$\tilde{V}_{Az}=(\langle |B_z|^2\rangle_{\phi,z}/\langle \rho\rangle_{\phi,z})^{1/2}$ from the last $100T_\text{orb}$ of zn. Local analyses and semi-analytic calculations predict that in the presence of a strong toroidal magnetic field, r-mode frequency enhancements due to the additional restoring force of magnetic tension will be $\lesssim\omega_{Az}$ \citep[see section 3.2 in ][]{dew19}. Fig. \ref{fig:znAf375_pds} (middle) validates this prediction, showing the preferential excitation of magnetically altered inertial oscillations close in both frequency and radius to $\max[\kappa_{PW}+\omega_{Az}]$.

\begin{figure*}
    \centering
    \includegraphics[width=\columnwidth]{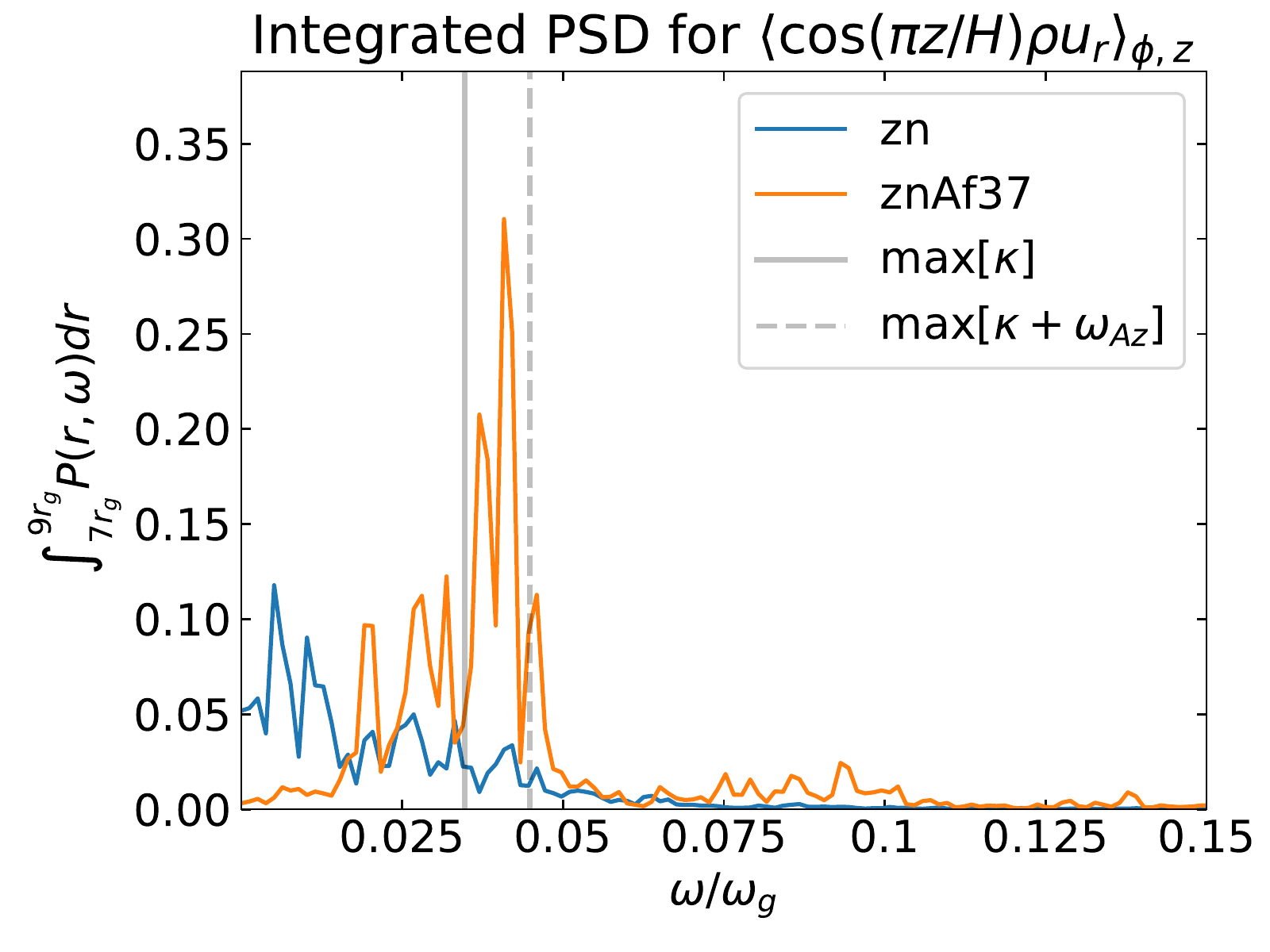}
    \includegraphics[width=\columnwidth]{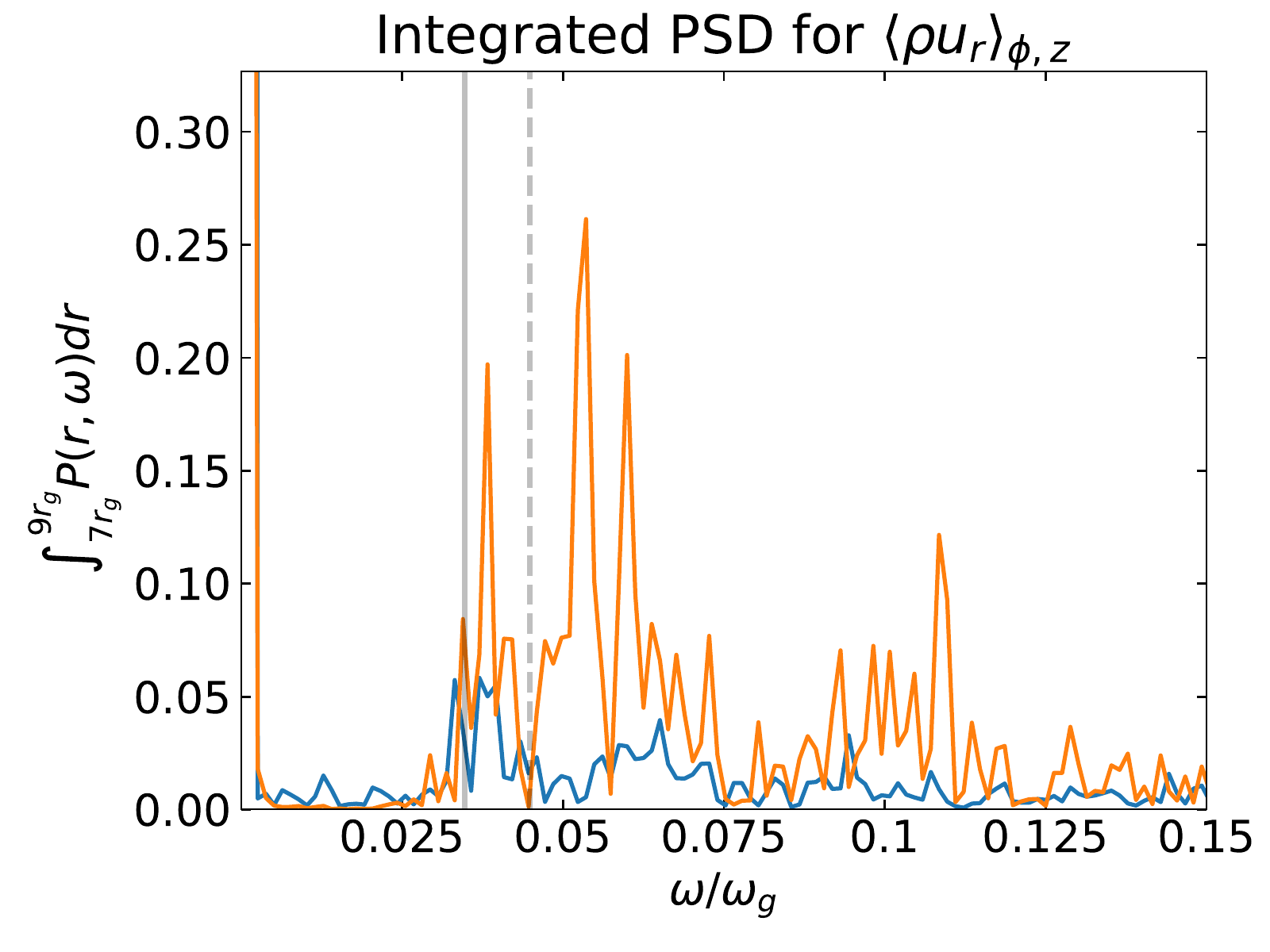}
    \caption{Plots comparing power spectral densities from simulations zn and znAf37, calculated by integrating the PSD $P(r,\omega)$ for $\langle\cos(\pi z/H)\rho u_r\rangle_{\phi,z}$ (left) and $\langle\rho u_r\rangle_{\phi,z}$ (right) (see Figs. \ref{fig:zn_pds} and \ref{fig:znAf375_pds}), over the radial domain $[7r_g,9r_g]$.  The PSDs are sampled over $t=120-200T_\text{orb}$ and normalized identically in each panel. The solid and dashed grey vertical lines indicate $\max[\kappa_{PW}]$ and $\max[\kappa_{PW}+\omega_{Az}]$. These 1D PSDs highlight an Alfv\'enic enhancement of trapped inertial wave frequencies (left), and enhanced f-mode power over a range of frequencies (right) in the eccentric disc simulation znAf37.} \label{fig:znAf_ipds}
\end{figure*}

The spacetime diagram and PSD for $\langle\rho u_r\rangle_{\phi,z}$ shown in Fig. \ref{fig:znAf375_spct} (bottom) and Fig. \ref{fig:znAf375_pds} (right) again illustrate the radial propagation of axisymmetric inertial-acoustic waves (f-modes), but with greater amplitudes and power than in the analogous timing data for the circular disc simulation zn (compare against Fig. \ref{fig:zn_spct}, bottom and Fig. \ref{fig:zn_pds}, right). In Paper I we found that a non-linear self-coupling of trapped inertial oscillations can produce axisymmetric f-modes at twice the r-modes' frequency. The axisymmetric f-modes excited in znAf37 do not produce near-integer ratios as clearly as in Paper I, though, the amplified inertial-acoustic waves' power extending over a range of frequencies both comparable to and greater than those of the (magnetically enhanced) trapped inertial waves.

Fig. \ref{fig:znAf_ipds} makes a closer comparison between the circular and eccentric disc simulations zn and znAf37, plotting radially integrated PSDs for $\langle \cos(\pi z/H)\rho u_r\rangle_{\phi,z}$ (left) and $\langle \rho u_r\rangle_{\phi,z}$ (right) from the two runs. Simulation znAf37 (orange line) shows enhanced variability in both PSDs. The variability in $\langle \cos(\pi z/H)\rho u_r\rangle_{\phi,z}$ illustrates the preferential excitation of (magnetically enhanced) trapped inertial waves. The signal is clearly only quasi-periodic, showing power enhancement over a range of frequencies including $\max[\kappa_{PW}]$ (solid grey  line) and $\max[\kappa_{PW}+\omega_{Az}]$ (dashed line), rather than the isolated peaks exhibited by our more controlled hydrodynamic simulations (cf., fig. 9, right in Paper I). This may be due in part to the dynamical nature of the background flow: while idealised, coherent magnetic fields produce smooth frequency changes in linear r-modes, we do not expect turbulent field fluctuations to provide a time-independent restoring force \citep{dew18}. From a numerical standpoint, without finely tuned initial conditions or the rigid lid employed in Paper I, we also do not expect the formation of coherent standing wave oscillations in our unstratified, cylindrical framework.

The integrated PSDs for $\langle \rho u_r\rangle_{\phi,z}$ (Fig. \ref{fig:znAf_ipds}, right) further indicate that significantly more inertial-acoustic waves are excited in the eccentric disc simulation znAf37 than in zn. Enhanced f-mode variability at frequencies greater than those of the trapped inertial waves suggests non-linear coupling, while amplified peaks falling in the inertial range instead indicate mode \emph{mixing}.
\footnote{
Mixing between vertically structured r-modes and vertically un-structured f-modes can occur due to a breaking of symmetry with respect to the mid-plane by mean magnetic fields with both vertical and toroidal components \citep{dew19}.
}
However, the f-mode spectrum covers too broad a frequency range to identify a frequency commensurability with any confidence.

Lastly, the enhanced Maxwell stress seen in zn2h motivates discussion of trapped inertial wave excitation in simulations with a larger vertical extent, although the cylindrical approximation is formally invalid on vertical scales $\gtrsim H$. Simulation zn2hAf37 was initialized from the saturated state of zn2h and similarly covers the vertical extent $z\in[-2H,2H].$ PSDs from the simulation show similar enhancements in variability to those exhibited by znAf37. The Maxwell stress in zn2hAf37 decreases at the same rate as in znAf37 (see below), but, starting from a higher initial value, remains larger at the end of run-time (see Table \ref{tab:trExM}).

\vspace{1em}
{\flushleft\textbf{Regime III: MRI suppression}}
\vspace{0.5em}

\begin{figure}
    \centering
    \includegraphics[width=\columnwidth]{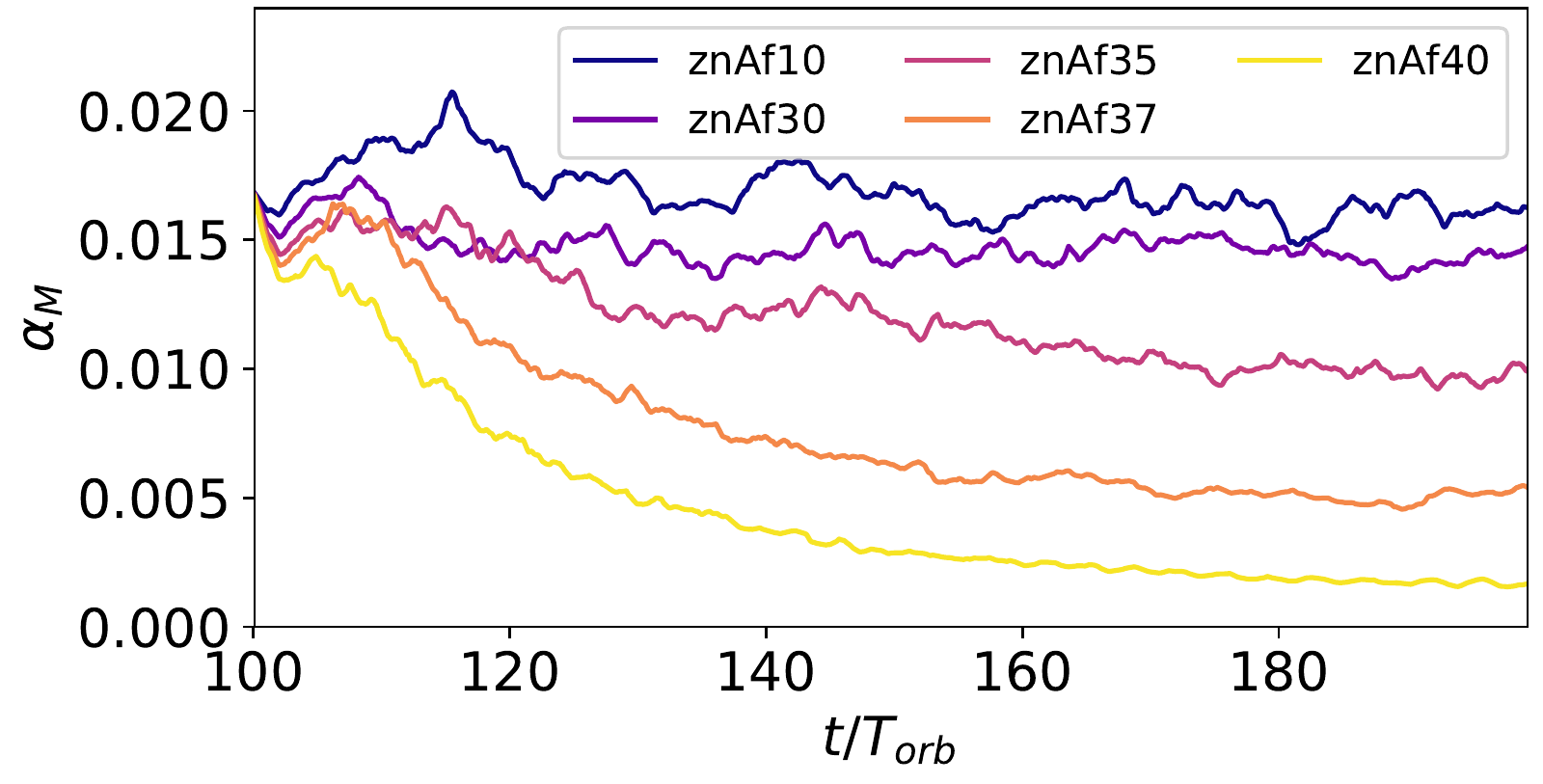}
    \caption{Plot showing the evolution of the Maxwell stress in simulations znAf10-40. The decreases in $\alpha_M$ over time illustrate the weakening of MHD turbulence by eccentric disc distortion.}
    \label{fig:AfMax}
\end{figure}

In this section we describe the weakening and even suppression of the MRI turbulence by increasingly strong disc distortions, best exhibited by the simulation znAf40. Qualitatively, the sequence of colour-plots in Fig. \ref{fig:streamax} (top) illustrate a laminarization of the flow, while the bottom colour-plots show a corresponding suppression of the Maxwell stress. Quantitatively, the eccentric distortions' impact on the MRI can be tracked most visibly via the Maxwell stress
$\alpha_M=-\langle B_rB_\phi\rangle_V/\langle P\rangle_V,$ which we plot in Fig. \ref{fig:AfMax}. The figure clearly shows a decrease in $\alpha_M$ that becomes more severe with increasing eccentricity forcing amplitude. While simulations znAf10-37 maintain modest but finite values of $\alpha_M,$ in znAf40 the disc distortion nearly completely suppresses the MHD turbulence by the end of the run-time.

In the absence of MRI turbulence in the inner disc, znAf40 forms the laminar `elevator flows' described in Paper I. These elevator flows, which are an artifact of periodic vertical boundary conditions in the cylindrical model, halt the growth of trapped inertial waves. They can be eliminated by an artificially rigid vertical boundary condition, which we have chosen to avoid in order to mitigate any spurious numerical effects on the MHD turbulence. Alternatively, the inclusion of vertical gravity and density stratification in future work should naturally eliminate the elevator flows. Their appearance in fact highlights the similarity of oscillations excited in znAf40 to those simulated in Paper I, and suggests that in some cases the suppression of the MRI by disc distortion might allow hydrodynamic instabilities (in particular r-mode excitation) to proceed unimpeded.

\section{Discussion}\label{sec:disc}
In this section we discuss the observational relevance of the conditions for the trapped inertial wave excitation demonstrated by our simulations, potential explanations for the weakening and suppression of MRI turbulence in our more strongly eccentric runs, and directions for future work.

\subsection{Trapped r-mode excitation and HFQPOs}
The simulations described in Section \ref{sec:ecc} demonstrate that a sufficiently strong disc eccentricity can enhance variability in the inner regions of relativistic, MHD-turbulent accretion discs. Our timing analysis associates this enhanced variability with Alfv\'enic-epicyclic inertial waves, excited and radially confined near the maximum in $\kappa.$ This demonstration of trapped inertial wave excitation in the presence of MRI turbulence suggests that magnetic fields may not be as destructive to r-modes as predicted by \citet{FL09} and \citet{ReM09}. Our simulations therefore provide support for the discoseismic explanation for HFQPOs, but with important caveats.

For one thing, we find that in the presence of the MRI turbulence, trapped inertial wave excitation requires a larger eccentricity ($e\gtrsim0.03$) near the ISCO than in the laminar, hydrodynamic simulations presented in Paper I of this series, suggesting that MRI turbulence actively damps the trapped modes. This value of eccentricity is likely a lower limit, since the zero-net-flux vertical magnetic fields initialized in our simulations produce only moderate levels of turbulence ($\alpha\sim0.03-0.04$). 

As detailed in Section 2.3 of Paper I, eccentric distortions excited at a 3:1 resonance with the orbit of the X-ray binary \citep[e.g.,][]{lub91a} can very reasonably be expected to propagate all the way to the inner disc, especially in the presence of elevated accretion rates associated with the SPL state \citep[see fig. 5 in ][or fig. 2 in Paper I]{fer09}. Observations of superhumps in BHBs \citep{odo96,nei07,zur08,kos18} indicate that black hole accretion discs in many of the systems are likely to be eccentric, although actual eccentricities are difficult to estimate, and to our knowledge there have been no direct measurements. 

On the other hand, decreases in eccentricity within $\sim20r_g$ due to non-linear effects, predicted by \citet{lyn19} and exhibited by both our MHD and hydrodynamic (see Appendix \ref{sec:2Dh}) simulations, may place a limit on eccentricities near the r-mode trapping region. In discs sustaining more vigorous turbulence than that produced by our zero-net-flux, non-radiative simulations, this upper limit on eccentricity may not be enough to overcome turbulent r-mode damping. On yet another hand, the non-linear steepening of eccentric waves could provide a more favorable environment for the distortions' associated hydrodynamic instabilities (like r-mode excitation), by first weakening MRI turbulence.

Regardless, we reiterate that the non-linear coupling that excites trapped inertial waves in our simulations is not unique to eccentric distortions; warps provide an alternative route to excitation \citep{fer08}. Indeed, early simulations of `tilted'  discs exhibited enhanced variability \citep{hen09}, although in those particular runs the variability may not have been due to standing mode oscillations \citep{hen12}. Practical considerations motivated our focus on the timing properties of eccentric discs, since simulating warped discs requires the full inclusion of vertical gravity, at much greater numerical expense \citep[e.g.,][]{lis19}. But we expect the fundamental mechanism of r-mode excitation via a non-linear coupling with a non-axisymmetric distortion to translate from eccentric to warped accretion discs.

\subsection{MRI modification}\label{sec:dmri}
A complete analysis of the effects of disc distortions on MRI stability lies beyond the scope of this project. However, the distinct radial localization of the MRI-suppression in our simulations is suggestive and deserves some commentary. The Maxwell stress is quenched first near the ISCO, where the eccentricity amplitude is lower but the quasi-stationary eccentric waves steepen and nearly shock (see Fig. \mbox{\ref{fig:streamax}}, top); this points to the strong compression associated with the eccentric waves' non-linear steepening as a possible cause of the MRI's suppression.

The steepening of the eccentric deformations in our simulations can be described formally by enhancements of the eccentricity gradient (change in eccentricity with radial coordinate) and twist (change in longitude of pericentre with radial coordinate). The variation of eccentricity with radius in our simulations constitutes an important difference between our setup and that of \citet{chan18}, who found persistent instability in their local analysis of \emph{uniformly} eccentric discs with net vertical magnetic flux.

Although the eccentric distortions in our simulations are essentially stationary from a global standpoint, their non-axisymmetry means that orbiting fluid elements will suffer intense and periodic flow perturbations every orbit. Nearby fluid elements tethered by magnetic fields will therefore be sucked in and out of a weak shock on the MRI timescale. It is reasonable to expect that this interference would disrupt the classical instability mechanism of the MRI involving angular momentum transport via field lines tethering and pulling apart nearby fluid blobs. Indeed, the local analysis of \citet{chan18} suggests that even a uniform distribution of eccentricity (without significant compression) can decrease linear MRI growth rates. 

The non-linear steepening of eccentric waves may additionally alter the turbulent dissipation of the magnetic field; studies of resistive plasmas in a variety of contexts indicate that compression can enhance magnetic reconnection \citep[e.g.,][]{hes11,uzd11}. There may also be a parallel with the suppression of zero-net-flux MRI by gravitoturbulence in vertically stratified shearing boxes \citep{rio18}, since the gravitational instability proceeds via spiral density waves with similarly strong density variations.

While the impact of distortions on the MRI may be less significant far away from a black hole, where eccentric modes have much longer wavelengths, compression due to strong eccentricity gradients may still affect MHD turbulence in a wider variety of astrophysical environments. The role played by eccentricities in MHD accretion discs should therefore be explored further in more controlled (i.e., non-relativistic) numerical simulations. Eccentric \citep{ogi14,wei18} and/or warped \citep{tor00,ogi13,paa19} shearing boxes likely offer the best approach to exploring the effects of disc distortion on the MRI.

\subsection{Future work}
Along with further investigation into the effects of disc distortion on MRI turbulence, several other avenues should be explored so as to better evaluate the viability of the discoseismic model for HFQPOs. While our simulations demonstrate that sufficient eccentricity can excite trapped inertial waves despite damping by MRI turbulence, we have excluded relevant physics in an effort to simplify a messy problem. For one thing, in the absence of vertical gravity or the rigid vertical boundary condition used in Paper I, the trapped inertial waves excited in our MHD simulations do not form global standing modes in $z$. Future work would likely require the full inclusion of density stratification to realize the formation of bona fide global modes.

Along with vertical gravity, future investigations of the discoseismic model for HFQPOs in X-ray binaries should include more detailed radiative and thermal physics relevant to the elevated states of accretion in which HFQPOs are observed. The connection between disc oscillations and hot coronal plasma (which produces the emission on which HFQPOs are imprinted) should also be an essential component of any comprehensive model. Coherent magnetic fields might provide such a connection by facilitating a transfer of variability from oscillations in the disc to time-dependent reconnection and emission in the corona, but exploring this possibility will require more sophisticated simulations \citep[perhaps drawing inspiration from studies of wave-interactions in the solar corona; e.g., ][]{pot19,mor19,rei19}.

\section{Conclusions}\label{sec:conc}
This paper describes the results of global, magnetohydrodynamic simulations of relativistic, eccentric accretion discs. These simulations show that trapped inertial waves (r-modes) can grow in the presence of MHD turbulence if excited by sufficiently strong disc distortion. These oscillations are important as they may drive high-frequency, quasi-periodic oscillations seen in the emission from black hole X-ray binaries. The trapped inertial wave growth is accompanied by outwardly propagating, axisymmetric inertial-acoustic waves (f-modes). Additionally, our simulations reveal a disruption of the magnetorotational instability in the inner regions of sufficiently distorted discs, likely owing to strong, non-axisymmetric compression associated with steepening and circularisation of the eccentric waves.

Like \cite{ReM09}, we do not observe signatures of trapped r-modes in simulations of \emph{circular} discs. However, in the presence of disc eccentricity, which we force into the domain with a non-axisymmetric density profile imposed at the outer boundary, we find three dynamical regimes. Simulations with low-amplitude forcing for eccentricity do not exhibit signatures of trapped inertial waves, likely due to r-mode damping by turbulent fluctuations. In runs in which larger forcing amplitudes produce sufficient eccentricity close to the ISCO, on the other hand, we do observe signatures of trapped inertial wave excitation in the power spectral density, along with a weakening of the Maxwell stresses associated with MRI turbulence. For eccentricities large enough that the MRI is completely suppressed and the inner disc is essentially laminar, we observe dynamics similar to the hydrodynamic simulations presented in \cite{dew20a} (Paper I).

Our simulations are simplified (isothermal, unstratified, zero-net-flux), and validation of the discoseismic model for HFQPOs will require the inclusion of more sophisticated physics. Still, this work demonstrates that sufficient disc distortion may, in principle, excite trapped inertial waves in black hole accretion discs despite damping by MHD turbulence.

\section*{Acknowledgements}
The authors thank the anonymous reviewer for comments and suggestions, which improved the quality of the paper. J. Dewberry thanks Adrian Barker, Roman Rafikov, Omer Blaes and Dong Lai for very helpful discussions. This work was funded by the Cambridge Commonwealth, European and International Trust, the Vassar College De Golier Trust, the Cambridge Philosophical Society, and the Tsung-Dao Lee Institute.  

\section*{Data availability}
The data underlying this article will be shared on reasonable request to the corresponding author.


\bibliographystyle{mnras}
\interlinepenalty=10000
\bibliography{trExMHDsim} 


\appendix

\newpage
\section{Hydrodynamic test simulations}\label{sec:2Dh}

\begin{table}
\centering
 \caption{Table listing (i) simulation label, 
 (ii) eccentricity forcing amplitude,
 (ii) saturated Reynolds stress, 
 (iv) radial KE,
 (v)  $\max|\tilde{E}|$ within $18r_g$,
 and (vi) $\max|\tilde{E}|$ between $6-10r_g$. 
The values have been averaged over $30-50T_\text{orb}$ for hAf10-40, and $100-150T_\text{orb}$ for hlAf40.}\label{tab:hydro}
\begin{tabular}{lcccccr} 
    \hline
    Label & 
    $A_f$ & 
    $\alpha_R$ &
    $\langle \rho u_r^2\rangle_V/\langle P\rangle_V$ & 
    $\max|\tilde{E}|$ & 
    $\max_D|\tilde{E}|$ \\
    \hline
    \hline
    hAf10  & $0.10$  & 0.0002 & 0.0089 & 0.011 & 0.010 \\
    hAf30  & $0.30$  & 0.0070 & 0.1298 & 0.043 & 0.027 \\
    hAf35  & $0.35$  & 0.0123 & 0.2217 & 0.057 & 0.031 \\
    hAf37  & $0.375$ & 0.0164 & 0.2933 & 0.065 & 0.033 \\
    hAf40  & $0.40$  & 0.0222 & 0.3944 & 0.075 & 0.034 \\
    hlAf40 & $0.40$  & 0.0107 & 0.1643 & 0.088 & 0.037 \\
    \hline
 \end{tabular}
\end{table}

In this appendix we describe 2D, purely hydrodynamic simulations (summarised in Table \ref{tab:hydro}) that provide a point of direct comparison for the eccentric distortions driven in our MHD simulations. These runs are motivated primarily by an inward decrease of eccentricity amplitude not seen in the much more weakly forced deformations in Paper I. In exhibiting very similar distortions to the MHD simulations described throughout this paper, the simulations presented in this appendix strongly indicate that the inward decrease in eccentricity is caused primarily by steepening and circularisation of the eccentric waves due to non-linear, purely hydrodynamic effects, rather than turbulent damping.

Simulations hAf10-40 use exactly the same sound speed, horizontal resolution, and radial boundary conditions as znAf10-40. Meanwhile, hlAf40 has been run on the much larger radial grid $r/r_g\in[4,54]$, with a resolution of $N_r\times N_\phi=744\times800$. To generate an eccentricity-forcing boundary condition for this simulation, we take the ratio of outer to inner boundaries as $r_1/r_\text{ISCO}=9$ in calculating an eigenmode of the non-linear theory considered by \citet{bar16} (see Section \ref{sec:BC}, or Paper I). Note that this calculation produces a weaker forcing than that used in hAf40, which has $r_1/r_\text{ISCO}=3$, since a larger radial domain implies that a shallower eccentricity gradient is required to produce the same maximum eccentricity in the two (Newtonian) calculations.

\begin{figure}
    \centering
    \includegraphics[width=\columnwidth]{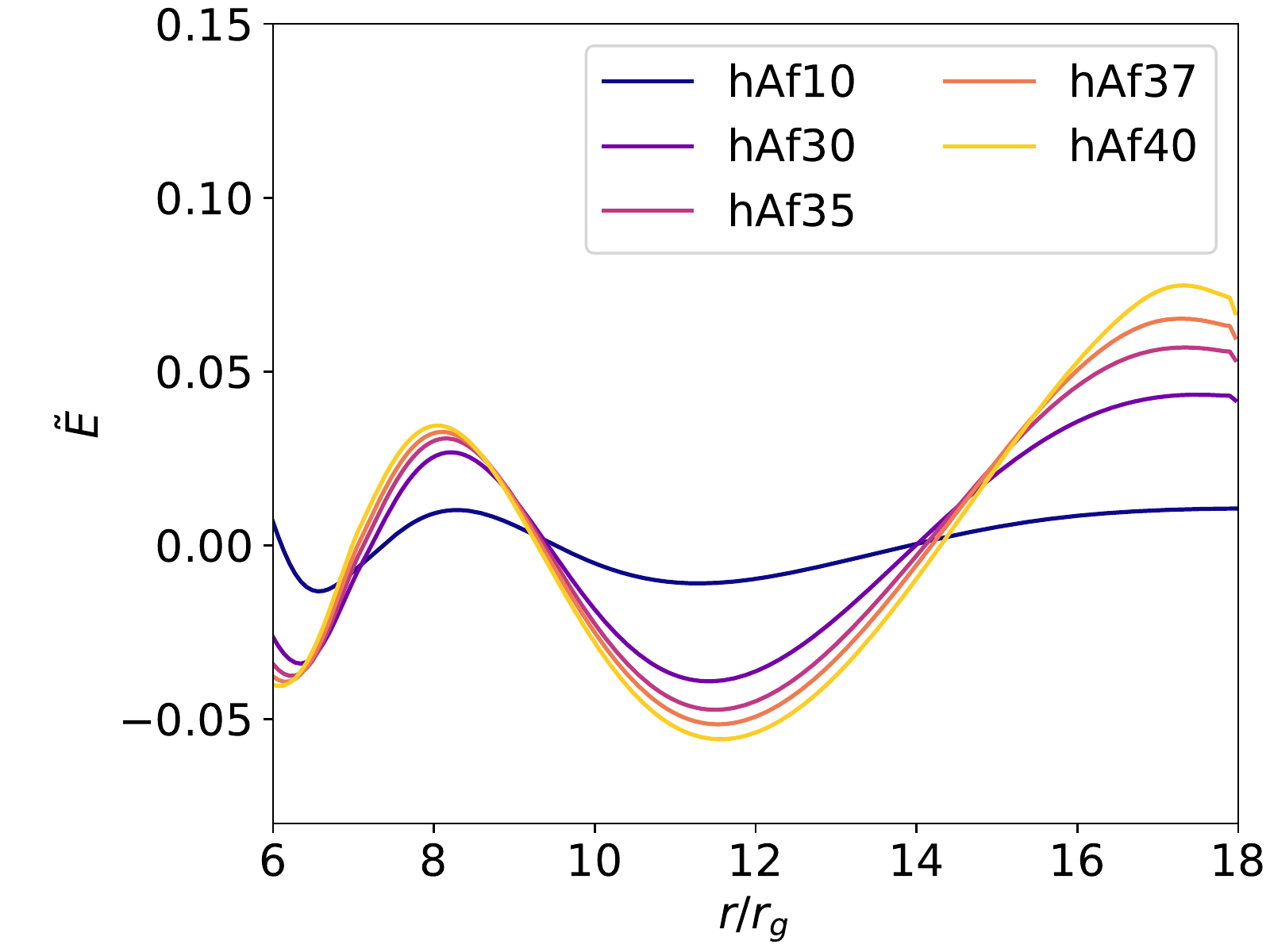}
    \caption{Same as Fig. \ref{fig:ecc}, but for the 2D, hydrodynamic simulations hAf10-40. The plot illustrates similar profiles of $\tilde{E}$ to the MHD runs znAf10-40. }\label{fig:hecc}
\end{figure}

\begin{figure*}
    \centering
    \includegraphics[width=\textwidth]{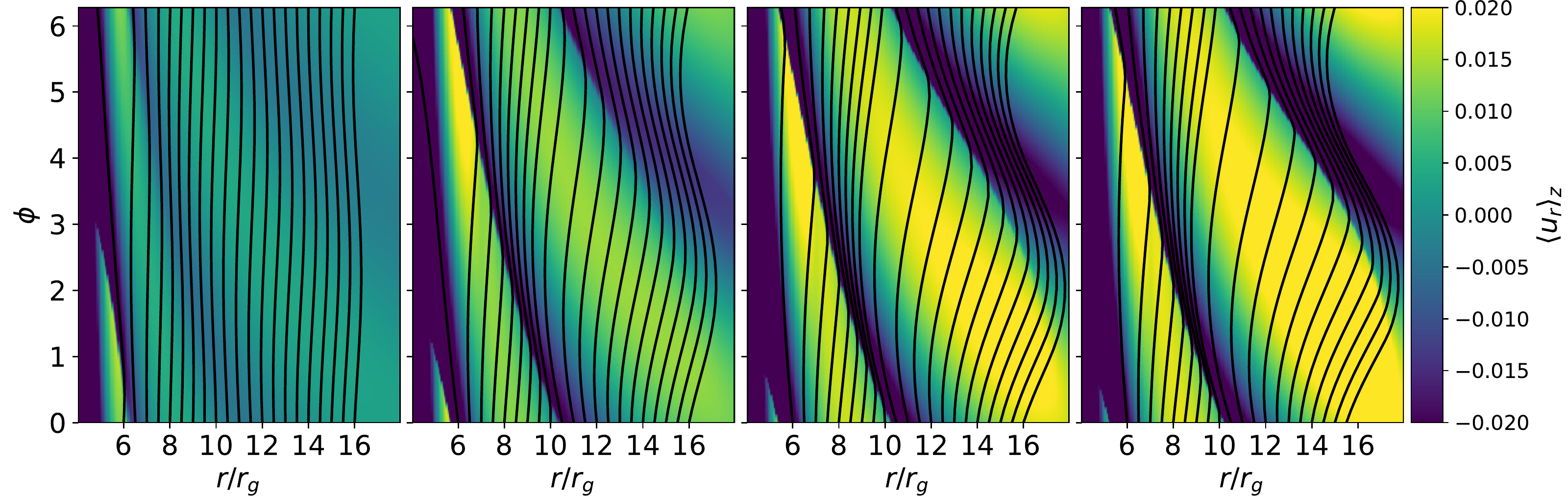}
    \caption{Same as Fig. \mbox{\ref{fig:streamax}} (top), but for the 2D, hydrodynamic simulations hAf10-40. The snapshots show very similar distortions to those in znAf10-40.}\label{fig:hstream}
\end{figure*}

The hydrodynamic simulations hAf10-40 produce very similar distortions to their MHD counterparts znAf10-40. The radial profiles of $\tilde{E}$ plotted in Fig. \ref{fig:hecc} are understandably smoother than, but still remarkably similar to, those shown in Fig. \ref{fig:ecc}. Fig. \ref{fig:hstream} further illustrates the similarity, displaying very similar colour plots to those shown in Fig. \ref{fig:streamax} (top). Comparing Table \ref{tab:hydro} with Table \ref{tab:trExM}, the hydrodynamic simulations show values of $\tilde{E}$ that are smaller by $\lesssim10\%$ than the corresponding MHD simulations in the outer disc (potentially due to an absence of turbulent velocity fluctuations), and larger by $\lesssim5\%$ near the trapping region. This indicates that the MHD turbulence damps the eccentricity amplitudes by no more than $\sim10-15\%$. The saturated values of $\alpha_R$ and $\langle \rho u_r^2\rangle_V/\langle P\rangle_V$ listed in Table \ref{tab:hydro} are also similar to the difference between the circular and eccentric MHD simulations zn and znAf10-40, although these quantities are not strictly comparable because of the MRI weakening that takes place in the latter simulations.

\begin{figure*}
    \centering  
    \includegraphics[width=\textwidth]{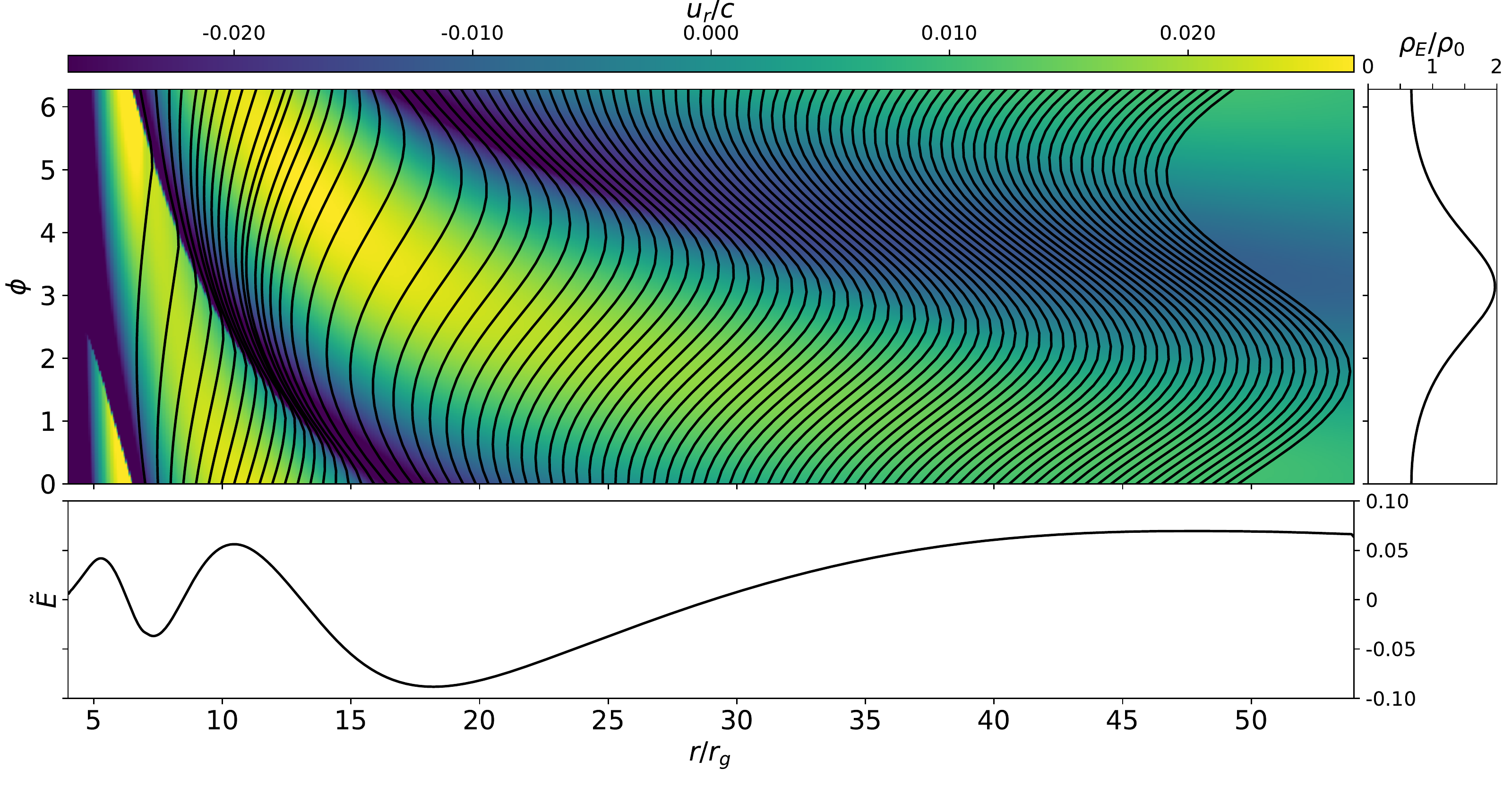}
    \caption{Top left: colour-plot showing radial velocity overlaid by streamlines in hlAf40. Top right: azimuthal profile of the density profile enforced in the outer radial ghost cells. Bottom: plot of $\tilde{E}\sim -e\sin\varpi$, illustrating that the eccentric wave circularisation occurs primarily within $r\lesssim15r_g$.}\label{fig:hlAf40}
\end{figure*}

Lastly, Fig. \ref{fig:hlAf40} describes the eccentric distortion forced in the larger domain run hlAf40, showing an unfolded polar snapshot of radial velocity overlaid by streamlines (top left), the density profile enforced in the outer ghost cells (top right), and a radial plot of $\tilde{E}$ (bottom). The figure demonstrates first that the much weaker forcing at large radii still gives rise to the values of $\tilde{E}\sim0.05-0.010$ observed for hAf40 and znAf40. Secondly, the plots in Fig. \ref{fig:hlAf40} show that the wave-steepening observed in our other simulations is not unique to the shorter radial domain, illustrating an eccentric distortion with a much larger wavelength at larger radii. Importantly, this rapid increase in wavelength with radius implies that turbulent eccentricity damping will be even weaker throughout most of the disc. Eccentric distortions should therefore face little difficulty in propagating to the ISCO, especially in the presence of the elevated mass accretion rates associated with the SPL state.

\bsp	
\label{lastpage}
\end{document}